%% file: report.tex
\documentclass[10pt, report,onecolumn]{report}
\usepackage{array}
\usepackage{color}
\usepackage{ifpdf}
\usepackage{graphicx}
\usepackage{epsfig} 
\usepackage[caption=false,font=footnotesize]{subfig}
\newcommand{\com}[1]{\textcolor{red}{#1}}
\usepackage{amsmath}
\usepackage{cite}
\usepackage[table]{xcolor}
\usepackage{url}
\usepackage{soul}
\usepackage{titlesec}
\usepackage[utf8]{inputenc}
\titleformat{\chapter}[display]
{\normalfont\bfseries\filcenter}
{\LARGE\thechapter}
{1ex}
{\titlerule[2pt]
\vspace{2ex}%
\LARGE}
[\vspace{1ex}%
{\titlerule[2pt]}]
\usepackage[letterpaper, total={6in, 9.5in}]{geometry} 
\newcommand{\figref}[1]{Fig.~\ref{fig:#1}}
\newcommand{\tblref}[1]{Table~\ref{tbl:#1}}
\newcommand{\secref}[1]{Section~\ref{sec:#1}}

\begin{document}

\input{figs}

\input{tabs}

\begin{titlepage}
	\centering
	\vspace*{4cm}
	{\scshape\Huge MMDF2018 Workshop Report \par}
	\vspace{2cm}
	{\Large\itshape Chun-An~Chou, Xiaoning~Jin, Amy~Mueller, and~Sarah~Ostadabbas \par}
	\vspace{2cm}
	{\Large \today\par}
	\vfill
    \includegraphics[width=\linewidth,trim=0in 0in 0in 0in, clip=true]{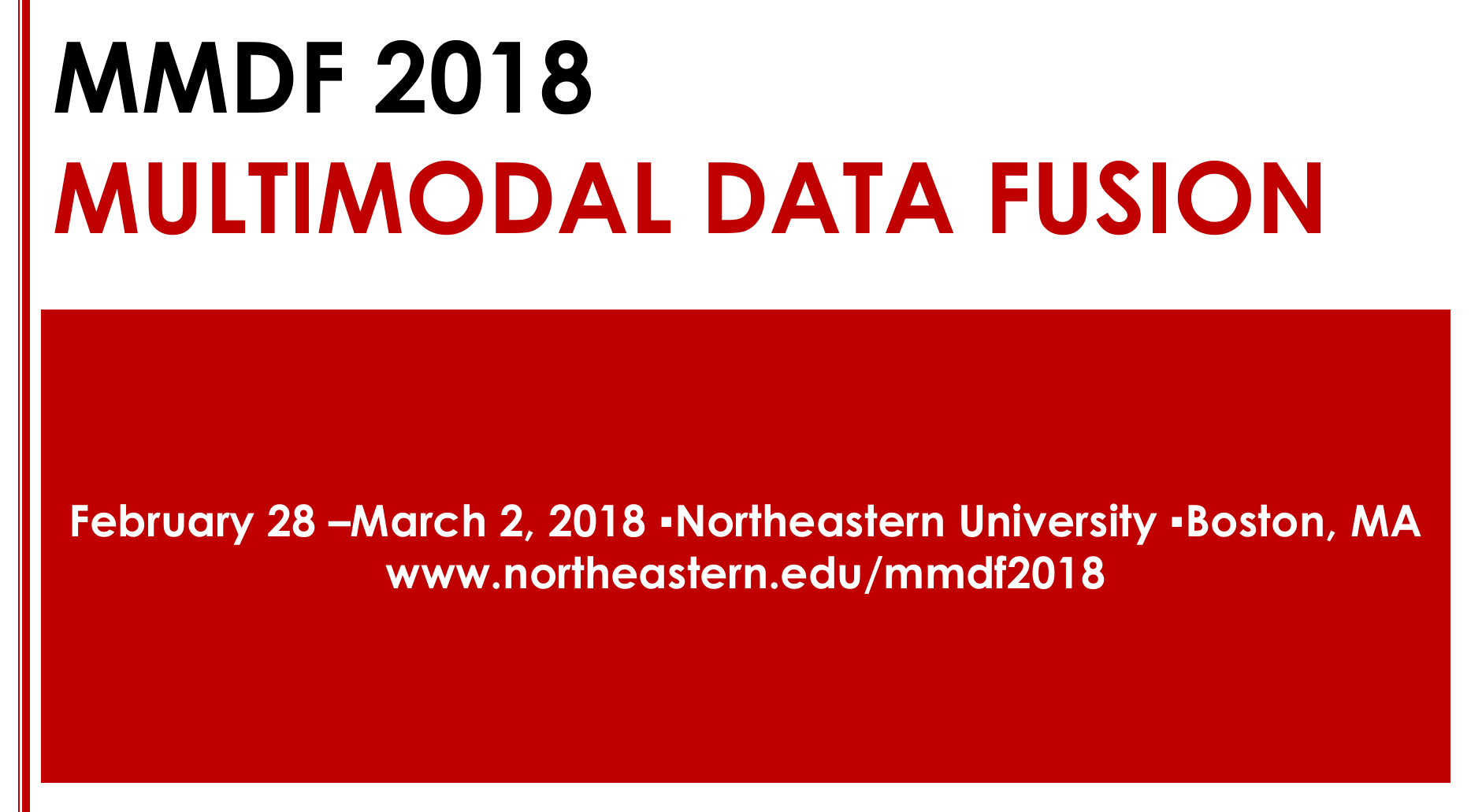}
\end{titlepage}

\chapter*{MMDF2018 Committees and Planning}

\section*{\centering{\textbf{Organizing Committee}}}
\begin{tabular} {ll}
\hline
Prof. Chun-An Chou &    Mechanical and Industrial Engineering, Northeastern University \\
Prof. Xiaoning Jin &    Mechanical and Industrial Engineering, Northeastern University \\
Prof. Amy Mueller &     Civil and Environmental Engineering, Northeastern University \\
Prof. Sarah Ostadabbas &    Electrical and Computer Engineering, Northeastern University \\

\hline
\end{tabular}

\section*{\centering{\textbf{Advisory Committee}}}
\begin{tabular} {ll}
\hline

Prof. Akram Alshawabkeh	&   Civil and Environmental Engineering, Northeastern University \\
Prof. Lisa Feldman Barrett &   Psychology, Northeastern University \\
Claire Duggan	        &   Center for STEM Education, Northeastern University \\
Prof. Deniz Erdoğmuş 	&   Electrical and Computer Engineering, Northeastern University \\
Prof. Sagar Kamarthi 	&   Mechanical and Industrial Engineering, Northeastern University \\
Prof. Haris Koutsopoulos&   Civil and Environmental Engineering, Northeastern University \\
Prof. Stacy Marsella 	&   Computer Science, joint with Psychology, Northeastern University \\
Prof. Mark Patterson	&   Marine and Environmental Science, Northeastern University \\
Dr. Leonard Polizzotto	&   Executive in Residence, Northeastern University \\
Prof. Carey Rappaport	&   Electrical and Computer Engineering, Northeastern University \\
Prof. Hanumant Singh 	&   Electrical and Computer Engineering, Northeastern University \\
Prof. Milica Stojanovic &   Electrical and Computer Engineering, Northeastern University \\
Dr. Tommy Thomas		&   Neurology, Emory University School of Medicine \\
\hline
\end{tabular}

\vspace{20mm} 

\section*{\centering{\textbf{Acknowledgements}}}
The Organizing Committee would like to recognize and sincerely thank the Northeastern University university academic leadership, Office of the Provost (Provost and Professor James C. Bean and Vice-Provost for Research and Professor Arthur F. Kramer), and College of Engineering (and the Dean of Engineering, Professor Nadine Aubry) for providing invaluable support with funds, organization, and logistics.  We would also like to express our sincere gratitude to the members of the Advisory Committee, whose input helped shape this event and bring it into existence, to the Logistics Chairman Gaby Fiorenza for making sure the event came together, and to the dedicated team of student volunteers who executed her plan.  Finally we would like to thank ThermoFisher Scientific for their  sponsorship of the workshop Welcoming Reception and Dinner.

\pagebreak

\tableofcontents
\newpage

\chapter{Introduction}\label{sec:introduction}

\section{Motivation for the MMDF2018 Workshop}
Driven by the recent advances in smart, miniaturized, and mass produced sensors, networked systems, and high-speed data communication and computing, the ability to collect and process larger volumes of higher veracity real-time data from a variety of modalities is expanding. However, despite research thrusts explored since the late 1990's \cite{hall1997introduction,manyika1995data}, to date no standard, generalizable solutions have emerged for effectively integrating and processing multimodal data, and consequently practitioners across a wide variety of disciplines must still follow a trial-and-error process to identify the optimum procedure for each individual application and data sources.  A deeper understanding of the utility and capabilities -- as well as the shortcomings and challenges -- of existing multimodal data fusion methods \emph{as a function of data and challenge characteristics} has the potential to deliver better data analysis tools across all sectors, therein enabling more efficient and effective automated manufacturing, patient care, infrastructure maintenance, environmental understanding, transportation networks, energy systems, etc. (see \figref{Overview}).  There is therefore an urgent need to identify the underlying patterns that can be used to determine \emph{a priori} which techniques will be most useful for any specific dataset or application.  This next stage of understanding and discovery -- i.e., the development of generalized solutions –- can only be achieved via a high level cross-disciplinary aggregation of learnings, and this workshop was proposed at an opportune time as many domains have already started exploring use of multimodal data fusion techniques in a wide range of application-specific contexts.

\figOverview

\section{Statement of Need}
Multimodal data fusion is already being used and explored in a number of highly varied disciplines (e.g., \cite{kunz2013map,sui2012review,germa2010vision,wu2004optimal,chatzis1999multimodal,jannin2000data,kannathal2007data,faundez2005data,dalla2015challenges,olson2012climate}).  A number of key challenges common across disciplines can be identified, summarized as follows: (1) How best to work with heterogeneous data at different (spatial or temporal) scales and resolutions, of different types, and with varying levels of reliability? (2) How best to deal with noisy, uncertain, missing, and conflicting data? (3) How best to enable the interaction of multiple modalities in the processing pipeline, and at what level (data, features, or decisions)? (4) How best to integrate datasets (fusion), and when fusing how best to evaluate reasonableness of assumptions, especially given the unknown nature of interactions across datasets? (5) How best to evaluate performance, i.e., how to define an objective function for varying degrees of knowledge about the underlying phenomena and which can take advantage of multiple modalities to improve the end result? 

While the commonality of the underlying mathematics has been widely discussed (e.g., \cite{lahat2015multimodal}), a major barrier to learning is the lack of common language or framework in this field;  because different domains have evolved multimodal data fusion techniques relatively independently, evaluations, results, and even descriptions of the techniques are reported in varying terminology that obscures what might otherwise be common ground.  A major contribution of this workshop, therefore, is the creation of a (preliminary) shared framework for discussion (and improvement) as well as a vision for a workshop format that overcomes this barrier to enable cross-disciplinary learning.

Previous workshops in related topic areas (Appendix, \secref{pastworkshops}) provided context for the current state of research as well as related ideas and concepts which informed this workshop design.  Such topic- and discipline-driven workshops cultivate precisely the experts who must come together to create bridges between the islanded topics.  In this spirit, MMDF2018 was specifically designed to facilitate interdisciplinary conversations, broaden views, build/strengthen collaborations across disciplinary lines, and provide a forum for comparing lessons learned from different application perspectives.

\section{Workshop Design and Objectives}
This workshop was envisioned to bring together experts (both data and application scientists and engineers) from a wide range of domains to drive the needed fundamental advances in the field of multimodal data fusion, providing a forum for learning that generalizes beyond individual applications and domains.  Motivating topics included cooperative development of a more unified language for multimodal data fusion and working toward domain-independent tools and techniques to facilitate extraction of knowledge from complex data to drive rapid solution advancement.

To achieve these goals, the Workshop Organizers (who collectively apply multimodal data fusion techniques to manufacturing, healthcare, human decision support, and environmental applications) assembled a highly interdisciplinary Advisory Committee with representation from computer science, psychology, neurology, transportation, robotics, system integration, subsurface sensing, communications, signal processing, marine science, and STEM education to ensure the workshop format and participants are as responsive as possible to the stakeholder communities. The participants were comprised of academic, industry, and national laboratory leaders, targeting individuals with deep experience in developing or using multimodal data fusion techniques for application in specific fields (a complete list of workshop participants is available in \secref{list}). A highly interactive, facilitated but discussion-driven breakout-and-reconvene format provided a forum for rapid aggregation of knowledge and learning.  Lessons recorded by graduate student scribes and session moderators have been aggregated into this report, with an aim to identify a comprehensive national agenda and roadmap for future research directions.  This report was distributed to participants for comment and contributions before publication.  
%
%

The MMDF workshop had four main objectives associated with the above mentioned goals:
\begin{itemize}
\item  \textbf{Examine the landscape of existing multimodal data fusion techniques across sectors:} Summarize the state-of-the-art research in multimodal data methodologies and techniques and applications where they are in use.  Identify the current needs of data fusion research across various sectors.  Identify the failures/shortcomings in existing techniques and lessons learned from experiences.
\item \textbf{Identify commonalities and differences, successes and gaps, in current techniques with respect to a range of dataset types:}
Identify multisensory/multimodal data fusion challenges.  Identify commonalities and differences in successful applications of data fusion techniques across various domains and applications (especially generalizable lessons).  Identify gaps in existing research and development, including persistent roadblocks, challenges, and barriers for future research. Determine resources required to conduct effective and efficient multimodal data fusion in high-impact application domains.
\item \textbf{Draft a national agenda and roadmap for multimodal data fusion research and development with the goal of enabling novel multimodal data fusion tools to propel progress:}
Identify the national and global drivers, attributes, imperatives, and barriers for multimodal data fusion methodologies and key application challenges. Formulate a shared multimodal data fusion framework to underpin development of tools that can be applied across disciplines to allow researchers to focus on lessons learned from data rather than trial-and-error data processing. Formulate recommendations for research programs and activities to advance multimodal data fusion solutions that bring together all involved communities. Formulate recommendations for integration of associated concepts into education at the K-12 (primarily high school), undergraduate, and graduate levels, to ensure a pipeline for an enthusiastic and capable future workforce in this field.
\item  \textbf{Disseminate the findings of this workshop widely across domains and sectors to reach all stakeholder communities, including research, industry, and education.}
\end{itemize}

\section{Workshop Discussion Framework}
As a starting point for discussion and a first step toward apples-to-apples comparisons, a preliminary (intentionally highly generalized) framework was provided to the participants, shown in \figref{Framework}. Two examples are provided in the Appendix \secref{examples}.
\figFramework

To understand the utility of \figref{Framework} in facilitating an ``apples-to-apples'' discussion consider two domain-specific examples: manufacturing line management and patient care (see \tblref{Applications}).  The types of data in these applications differ significantly (from images to time series to sentences), however ultimately in both applications at a minimum the user wants to know (1) how to optimize system efficiency/effectiveness and (2) how to evaluate uncertainty in results.  Despite the highly differing domains and challenges, the underlying mathematical structure can be mapped onto similar processes: highly diverse data streams must in some way be aggregated to be simultaneously fed to a processing algorithm, which itself provides the end-user with an estimate of the critical characteristics of interest for the system state.  However, what is the best way to do so?  Should one-dimensional features be extracted from each modality and those features examined together using some type of inference model?  Or should the full dataset be fed simultaneously into the inference model (and if so, how to transform 2D or 3D data?)?  In either scenario, which type of inference model will be best:  machine learning (e.g., artificial neural networks), statistical analysis (e.g., Bayesian maximum likelihood), fitting of parameters for a physics-based or empirical model, or something else?  How do we define the metric for which is best?  What role does expert domain knowledge play in each of these steps (feature extraction, inference, metric development)?  And, critically, once we determine the optimal methodology for application \#1, can it tell us anything about the optimal methodology for application \#2?  Answering this question is the goal of the burgeoning field of multimodal data fusion.

\section{Workshop Overview and Agenda}
The full workshop agenda is provided in the Appendix (\secref{agenda}), however the format and discussion topics are summarized briefly here. The workshop was held over two days, primarily centered around breakout-and-reconvene discussion sessions but with the first half-day comprised of plenary talks to frame the discussion.  Throughout all sessions the following perspectives were highlighted for deeper understanding and comparison: \newline

\textbf{(I) Applications \& Disciplines:}	\newline
\indent \indent $\diamond$ Healthcare and biomedicine (neuroscience, psychology, genomics, etc.) \newline
\indent \indent $\diamond$ Environmental science (geology, climate change, water quality monitoring, etc.) \newline
\indent \indent $\diamond$  Manufacturing and automation \newline
\indent \indent $\diamond$  Transportation and logistics \newline
\indent \indent $\diamond$ Energy (smart grid, smart agriculture, etc.) \newline
\indent \indent $\diamond$   Robotics \newline

\textbf{(II) Methodology \& Theory:}	\newline
\indent \indent $\diamond$  Supervised and unsupervised learning	\newline
\indent \indent $\diamond$  Transparent vs. black-box methods	\newline
\indent \indent $\diamond$  Probabilistic inference modeling	\newline
\indent \indent $\diamond$  Real-time vs. offline learning	\newline
\indent \indent $\diamond$  Anomaly detection	\newline
\indent \indent $\diamond$  Dimensionality reduction (feature extraction, feature selection, , etc.) 	\newline
\indent \indent $\diamond$  Levels of multimodal data fusion (raw data, feature, decision, etc.)	\newline

\pagebreak

\chapter{Plenary Session and Keynote}
The workshop was kicked off with a half-day plenary session that set the stage through presentation of key disciplinary challenges and existing technology solutions in data integration.  Presentations captured input from academia and industry, from both the data science and practitioner perspectives.

\section{Keynote:  Dr. Kenneth Loparo, Case Western Reserve University}
\textbf{Multimodal Data Analysis for Clinical Decision Support in Critical Care Medicine --} The Intensive Care Unit (ICU) is the most data intensive unit in the hospital.  This talk discussed streaming real-time multimodal patient data and the integration and analysis of this data to support clinical decision-making.

\textbf{Bio:} Dr. Kenneth A. Loparo was an assistant professor in the Mechanical Engineering Department at Cleveland State University from 1977 to 1979 and has been on the faculty of the Case School of Engineering at Case Western Reserve University since 1979.  He is the Nord Professor of Engineering in the Department of Electrical Engineering and Computer Science and holds academic appointments in the departments of biomedical engineering and mechanical and aerospace engineering in the Case School of Engineering.  He has received numerous awards including the Sigma Xi Research Award for contributions to stochastic control, the John S. Diekoff Award for Distinguished Graduate Teaching, the Tau Beta Pi Outstanding Engineering and Science Professor Award, the Undergraduate Teaching Excellence Award, the Carl F. Wittke Award for Distinguished Undergraduate Teaching and the Srinivasa P. Gutti Memorial Engineering Teaching Award.  He was associate dean of engineering from 1994--1997 and chair of the Department of Systems Engineering from 1990--1994.  Dr. Loparo is a Life Fellow of the IEEE and a fellow of AIMBE, his research interests include stability and control of nonlinear and stochastic systems with applications to large-scale systems; nonlinear filtering with applications to monitoring, fault detection, diagnosis, prognosis and reconfigurable control; information theory aspects of stochastic and quantized systems with applications to adaptive and dual control and the design of distributed autonomous control systems; the development of advanced signal processing and data analytics for monitoring and tracking of physiological behavior in health and disease. 

\section{Dr. T{\"u}lay Adali, University of Maryland Baltimore County}
\textbf{Data Fusion through Matrix and Tensor Factorizations:  Uniqueness, Diversity, and Interpretability --}  Fusion of multiple sets of data, either of the same type as in multiset data or of different types and nature as in multi-modality data, is inherent to many problems in engineering and computer science. In data fusion, since most often very little is known about the relationship of the underlying processes that give rise to such data, it is desirable to minimize the modeling assumptions and, at the same time, to maximally exploit the interactions within and across the multiple sets of data. This is one of the reasons for the growing importance of data-driven methods in data fusion tasks. Models based on matrix or tensor decompositions allow data sets to remain in their most explanatory form while admitting a broad range of assumptions among their elements. This talk provided an overview of the main approaches that have been successfully applied for fusion of multiple datasets with a focus on the interrelated concepts of uniqueness, diversity, and interpretability. Diversity refers to any structural, numerical, or statistical inherent property or assumption on the data that contributes to the identifiability of the model, and for multiple datasets, provides the link among these datasets. Hence, diversity enables uniqueness, which is key to interpretability, the ability to attach a physical meaning to the final decomposition. The importance of these concepts as well as the challenges that remain are highlighted through a number of practical examples.

\textbf{Bio:} Professor T{\"u}lay Adali received the Ph.D. degree in Electrical Engineering from North Carolina State University, Raleigh, NC, USA, in 1992 and joined the faculty at the University of Maryland Baltimore County (UMBC), Baltimore, MD, the same year.  She is currently a Distinguished University Professor in the Department of Computer Science and Electrical Engineering at UMBC and is the director of the Machine Learning for Signal Processing Lab (MLSP Lab).  Prof. Adali is a Fellow of the IEEE and the AIMBE, a Fulbright Scholar, and an IEEE Signal Processing Society Distinguished Lecturer. She is the recipient of a 2013 University System of Maryland Regents' Award for Research, an NSF CAREER Award, and a number of paper awards including the 2010 IEEE Signal Processing Society Best Paper Award.  Her current research interests are in the areas of statistical signal processing, machine learning, and applications in medical image analysis and fusion.  Prof. Adali assisted in the organization of a number of international conferences and workshops including the IEEE International Conference on Acoustics, Speech, and Signal Processing (ICASSP), the IEEE International Workshop on Neural Networks for Signal Processing (NNSP), and the IEEE International Workshop on Machine Learning for Signal Processing (MLSP). Prof. Adali chaired the IEEE Signal Processing Society (SPS) MLSP Technical Committee (2003-–2005, 2011–2013), served on the SPS Conference Board (1998-–2006), IEEE SPS Signal Processing Theory and Methods (2010--2015) Technical Committee, and the IEEE SPS Bio Imaging and Signal Processing Technical Committee (2004-–2007). She was an Associate Editor for IEEE Transactions on Signal Processing (2003-–2006), IEEE Transactions on Biomedical Engineering (2007-–2013), IEEE Journal of Selected Areas in Signal Processing (2010--2013), and Elsevier Signal Processing Journal (2007-–2010). She is currently serving on the Editorial Boards of the Proceedings of the IEEE and Journal of Signal Processing Systems for Signal, Image, and Video Technology. 

\section{Dr. Alex Hall, Thermo Fisher Scientific}
\textbf{Effective Multi-Modal Multi-scale Analytical and Imaging Correlation --} Thermo Fisher Scientific builds and sells many different types of analytical imaging instruments and software.  Instruments include optical microscopes, Raman, EDX, XPS, electron microscopes, and micro-tomograph, each of which has particular strengths and weaknesses.  In order to answer more complex customer questions that require insight from multiple instrument data streams Thermo Fisher Scientific also develops software solutions, like Avizo and Amira, to correlate and fuse results from different instruments in more complex ways than `voting' with data from each independently.  This talk presented the main challenges faced in this emerging field and Thermo's ideas on how to answer them, supported by case study examples from current customer projects.

\textbf{Bio:} Dr. Hall is a Product Marketing Engineer representing the 3D analysis program Avizo-Amira with Thermo Fisher Scientific.  His background is in evolutionary biology where he used genetic data and skeletal variation collected from X-ray computed tomography to describe several species of reptiles.

\section{Dr. Timothy Heidel, National Rural Electric Cooperative Association}
\textbf{Leveraging Data to Enhance Power System Planning and Operations --}  Rapidly falling costs of distributed and renewable electricity generation and storage technologies and growing emphasis on improving electric power system resiliency have been motivating the investigation of alternative architectures for planning and operating electric power systems. Complementing these trends, advances in power electronics, computation, and communication technologies provide the opportunity to optimize and control grid operations closer to the locations where power is ultimately consumed. This could offer significant efficiency, cost, reliability, and emissions benefits. However, the methods that have been relied on for designing and operating power systems historically could prevent the full realization of the potential benefits associated with these technologies. Technologies that utilize data from a wide variety of sources across a variety of time scales will be critical to achieving the full potential of new technologies.  This talk focused on several examples where the use of new data streams may significantly enhance grid operations and planning.

\textbf{Bio:} Dr. Tim Heidel (now at Breakthrough Energy Ventures) gave this presentation as NRECA's Deputy Chief Scientist where he was leading R\&D activities, with emphasis in the areas of electric utility data analytics and grid cybersecurity.  Prior to joining NRECA, Dr. Heidel served as a Program Director at the Advanced Research Projects Agency-Energy (ARPA-E), where he managed a portfolio of over 75 research projects including new approaches for controlling and optimizing the transmission and delivery of electric power, particularly in the context of high renewables penetration.  As the Research Director for MIT's 2011 ``Future of the Electric Grid'' study, Dr. Heidel coordinated research efforts from economics, policy and electrical engineering on the most important challenges and opportunities that are likely to face the U.S. electric grid between now and 2030.  Dr. Heidel holds S.B., M.Eng., and Ph.D. degrees in Electrical Engineering and an M.S. in Technology and Policy from MIT.

\section{Dr. James Llinas, Center for Multisource Information Fusion, University at Buffalo}
\textbf{(Some) Remaining and Evolving Challenges to MMDF System Design and Development --} Much of the MMDF community remains focused on research and development of algorithms for narrow specific applications.  Such research remains needed but the community needs to lift its focus to more robust capabilities and more systemic issues related to achieving advanced capabilities.  This talk focused on some of what we call remaining challenges in MMDF, these focused on attaining both high-level MMDF capabilities as well as more fully integrated system architectures. The second category of challenges are labeled evolving, associated with what are in essence dramatic changes in the data and computational infrastructure, such as soft data and the Internet of Things. The summary message was an appeal for new initiatives across a broader R\&D base that involves focused multidisciplinary programs. This was joint work between Dr. James Llinas, University at Buffalo and Dr. Erik Blasch, Air Force Office of Scientific Research.

\textbf{Bio:} Dr. James Llinas is an Emeritus Professor in the Departments of Industrial and Systems Engineering and dual-appointed in the Department of Mechanical and Aerospace Engineering at the State University of New York at Buffalo.  He is the Founder and Executive Director of the Center for Multisource Information Fusion, the only systems-oriented academic research center for Information Fusion in the United States; the Center has been in existence for some 20+ years and has conducted extensive and distinctive research for a wide variety of governmental and civilian clients.  Dr. Llinas brings over 35 years of experience in multisource information processing and data fusion technology to his research, teaching, and business development activities.  He is an internationally-recognized expert in sensor, data, and information fusion, co-authored the first integrated book on Multisensor Data Fusion, and has lectured internationally for over 30 years on this topic.

\section{Dr. Soundar Kumara, Pennsylvania State University}
\textbf{Multimodal Sensing and Analytics--An Expository Look --} This talk examined the notion of multi modal sensing, and data analytics in the context of manufacturing and healthcare with a focus on the progression of sensor signal representation in manufacturing and how smart manufacturing and sensing has evolved in the past three decades.  Use cases from healthcare and IoT were discussed.

\textbf{Bio:} Dr. Soundar Kumara is the Allen, E., and Allen, M., Pearce Professor of Industrial Engineering at Penn State.  Has an affiliate appointment with the school of Information Sciences and Technology.  His research interests are in Sensor based Manufacturing Process Monitoring, IT in Manufacturing and Service Sectors, Health Analytics, Graph Analytics and Large Scale Complex Networks.  He got his undergraduate degree from S.V.U. College of Engineering, Tirupati, India; M.Tech., from I.I.T., Madras and Ph.D., from Purdue University.  Prior to attending Purdue he worked with the computing group of Indian Institute of Management, Ahmedabad.  He is a Fellow of Institute of Industrial Engineers (IIE), Fellow of the International Academy of Production Engineering (CIRP), Fellow of American Association for Advancement of Science (AAAS), and American Association of Mechanical Engineers (ASME).  He has won several awards including the Faculty Scholar Medal (highest research award at PSU), PSU Graduate Teaching Award, Penn State Engineering Society Outstanding Advisor Award, Premier Research Award, and PSES Outstanding Research Award.

\pagebreak
\chapter{Breakout Sessions}
The following section presents the MMDF Workshop results for the four main breakout sessions focused on the topics of healthcare, robotics/manufacturing, energy/transportation/environmental, and methodologies and theories.  For each breakout the discussions and activities are summarized. Participants (list provided in \secref{list}) identified goals, desired capabilities, challenges and barriers, and priority research topics.

\section{Track 1: Healthcare}

\subsection{Overview}
This breakout session, comprised primarily of researchers in academia and industry, discussed issues of MMDF primarily in acute/critical care and neuroscience.  As new medical sensing devices are developed to advance real-time data acquisition in ICU, the quality of acute/critical care can be potentially improved, however currently there is no gold standard to guide one to access, extract, and integrate information from highly-massive, inconsistent data of different modalities\cite{Johnson2016critical}. In neuroscience the issue of differing resolution and registration is significant:  among the most widely-used neuroimaging techniques to record brain activities, electroencephalogram (EEG) can generate high temporal resolution data at low signal-to-noise ratios whereas functional magnetic resonance imaging (fMRI) produces relatively high spatial resolution data \cite{sui2012review, lahat2015multimodal}.  Fusing data from these two modalities has shown a potential to uncover brain chaos and produce meaningful spatio-temporal information through intensive data analysis.  Therefore the overall goal of this breakout discussion was to discuss in detail: (1) accessibility, liability, usability, and interpretability of healthcare-related data, and (2) potential for generalized data fusion solutions to improve the clinical usage of data, i.e., more accurate and interpretable decisions upon defined research questions or patient conditions. 

\subsection{Successes}
With an existing awareness of MMDF tools and potential in this area, many successes using statistical/machine learning, signal/imaging processing, and information theoretic approaches have already been achieved in processing heterogeneous medical data.  In particular, machine learning has been employed to extract and fuse relevant attributes (features) from different data sources, leading to reliable prediction/inference outcomes in medical diagnosis, early detection, and prediction \cite{Johnson2016critical}. Thanks to advanced sensing technologies, more high-quality time-series data of vital signs, lab results, and 2D/3D images are available \cite{Pires2016datafusion}.  This may overcome some identified issues related to alignment and synchronization of multi-sensor time-series.  It may also enable a vision for an end-to-end learning framework (e.g., deep learning) for imaging and text recognition for various target diseases and disorders \cite{Purushotham2017deep,Mahmud2018deep,Suresh2018deep}.  Some recent efforts have been made to improve efficiency and interpretability of MMDF processes. 

\subsection{Challenges and Barriers}
Achieving the desirable capabilities and needs for MMDF requires overcoming several challenges and barriers, outlined in \tblref{HCChallenges}.

\tabHCChallenges

\subsection{Desired Capabilities and Priorities}
Based on the discussed barriers and challenges, desired capabilities are identified as outlined in \tblref{HCCapabilities}.

\tabHCCapabilities











\section{Track 2: Robotics/Manufacturing}

\subsection{Overview}
This breakout session focused on discussing the state-of-the-art science, methodologies and tools of MMDF in the domain areas of manufacturing, automation, and robotics.  As sensing and data acquisition technologies for machinery, automation and robotics continue to evolve, they provide richer and more relevant information for robust data-driven analytics for improved performance, safety, reliability and maintainability of robotic systems and manufacturing equipment or processes.  These emerging resources of multi-source multi-modal data collection and analytics, in conjunction with smart manufacturing and automation systems where cyber and physical systems are highly interconnected, are expected to improve the overall effectiveness, efficiency and fidelity of these systems. The goals identified for MMDF in the domain areas of robotics and manufacturing emphasize monitoring, detection, diagnostics, and substantial improvements in reliability (reduction in unexpected downtime), improved safety, operational efficiency, and human-machine collaboration effectiveness. These goals are categorized into three disparate classes in the short to medium time frames as to when they are anticipated to be developed.

\subsection{Successes}

Advancements in multi-sensor fusion have been developing rapidly over the last decade as the level of automation at manufacturing enterprises increases with the advent of new external sensor technologies which augment existing embedded sensors on the equipment.  Successes have been achieved in a number of areas, such as (1) real-time data representation for linear and nonlinear signals, (2) techniques for feature extraction and dimensionality reduction, (3) graph-based clustering to deal with large-scale data sets, (4) data representation schemes, and (5) sophisticated real-time data representation and visualization.  These techniques have been implemented in automated manufacturing processes and production lines in various industries such as automotive, semiconductor, aerospace, pharmaceutical and bio-medical products manufacturing with relatively high automation level, rich sensor data sources, and sophisticated data acquisition technologies.

\subsection{Challenges and Barriers}

Specific goals and desired functionality were identified for how MMDF can contribute to advancing the missions and objectives of many smart manufacturing and robotics initiatives. However, many of these cannot currently be achieved due to a number of existing challenges and barriers that remain under-addressed. \tblref{MfgChallenges} lays out a number of these scientific challenges and barriers for MMDF methodology development and its effective integration into modern manufacturing systems.

\tabMfgChallenges

\subsection{Desired Capabilities and Priorities}
Researchers continue to build and improve on the capabilities of existing multi-modal data fusion techniques and methodologies all the while exploring what new capabilities MMDF can perform to bring benefits to the smart manufacturing and automation industry as well as to robotics applications. A number of key desired capabilities and priorities were identified to support MMDF process, summarized in \tblref{MfgCapabilities} and detailed below. 

\tabMfgCapabilities

\begin{itemize}
    \item \textbf{Robotics:} A majority of robotics studies use multi-source sensor data from experiments and virtual-sensing data from simulations for purpose like environmental pattern recognition, robot condition monitoring, navigation, and motion control. Existing multimodal fusion algorithms are mostly designed for fusion at low level (data and feature), e.g., Gaussian mixture regression, Bayesian regression, Bayesian networks, multi-layer perception,deep learning, etc. Since robots generally operate in an uncertain and constantly changing environment, a synergistic operation of many modalities (thermal, visual, tactile, etc.) is desired for improved capabilities for real-time operation with high accuracy and robustness.
    
    \item \textbf{Manufacturing and Automation:} The rapid advancements of sensor technology, telemetric technologies, and computing technologies have resulted in rich data in both temporal and spatial spaces in automated manufacturing systems. With massive data readily available, there is an unprecedented increase in demand for advanced methodologies to unveil underlying relationships among events and to assist with various tactical and operational decisions (monitoring, detection, diagnosis, prognosis, control). Due to the complexity of manufacturing systems, enabling MMDF methodologies and algorithms of multimodal data fusion are desired to overcome data uncertainty, multi-resolution and heterogeneity mismatches, and the high dimensionality of the data, as well as the increasing requirements for decision-making capabilities. The priorities of MMDF for the next generation smart manufacturing and automated factory will be the development of a unified MMDF framework to integrate multiple data streams with disparate modalities with physical domain knowledge in manufacturing system modeling and analysis.
\end{itemize}

\section{Track 3: Energy/Transportation/Environmental}

\subsection{Overview}
This breakout session brought together researchers working broadly in infrastructure, security, and environmental areas to compare needs, approaches, successes, and challenges related to fusion of multimodal data.  Expertise areas covered included:  energy, transportation, climate, environment (ecology, fluid dynamics, chemistry), geotechnical, autonomous vehicles (underwater, transportation systems), and security (subsurface sensing, body scanning, etc.).  While diverse, these applications generally share characteristics of large system scale and significant system complexity, e.g., transportation or electrical distribution networks, global or local environments, thousands of feet of earth crust.

The main goals of MMDF applications for this class of systems are (1) understanding dynamics of complex systems at an appropriate and sufficient level to (2a) recommend architectures for optimization of real-time feedback controls (automated or human-in-the-loop) or (2b) optimize long-term resource management or usage policies under economic or other external constraints.  Models/algorithms are therefore required both to drive fundamental knowledge acquisition and to underpin decision support tools or automated management systems.  A main take-away from this session was the differential definition of `optimization' relative to traditional mathematical definitions; the majority of applications in this space require `cost-time constrained MMDF' and therefore should optimize based on a metric balancing output accuracy with economic costs and system response time.

\subsection{Successes}
To date successes in this area have primarily come through use of physics-, chemistry-, or other science-based models, including data assimilation techniques for guiding and verifying model accuracy.  This requires consistent gridding and synchronization, and resampling of data is common.  Significantly, the use of functional models provides an explicit pathway for knowledge capture, simulation, and interpretation of outputs in the context of system dynamics and other system properties.

While most applications are data limited, as indicated above, some data streams are sufficiently dense to enable estimation of extreme value distributions and, in more limited cases, the underlying probability distribution.  Probability theory for uncertainty and error propagation is relatively well-developed for these applications (facilitated by use of functional models).  Exploration of methods for creation of synthetic datasets - to overcome limitations in data availability for model training/calibration or to address privacy/security concerns - have led to some limited successes, e.g., the APRA-E GRID-DATA (Generating Realistic Information for the Development of Distribution and Transmission Algorithms) program.

\subsection{Challenges and Barriers}
In this breakout session challenges were identified at all stages of the MMDF process, from data characteristics to algorithm development to interpretability of results.  Particularly the latter is highlighted as a key consideration, given the goals of fundamental understanding of, and real-time management of, system dynamics, however issues related to diverse aspects of data sources comprised a significant fraction of the conversation, providing insight into opportunities for MMDF innovation (see \tblref{EnvChallenges}).

\tabEnvChallenges

\subsection{Desired Capabilities and Priorities}
Targets for MMDF capabilities and priorities were driven by the fundamental need in this sector for real-time controls on complex systems in environments where data collection is expensive, necessarily incomplete, and of highly varying accuracy (see \tblref{EnvCapabilities}).  Priorities were highlighted that deliver both an ability for operational decision making and for insight into underlying system dynamics (fundamental knowledge/intuition).  ``Confidence'' in data and results was a recurring theme, particularly as models/model outputs in these fields are often used by regulators and therefore must be justified in the context of fiscally responsible and/or safety-oriented management of public systems.

\tabEnvCapabilities

  \vspace{-.1in}
\section{Track 4: Methodologies and Theories}

\subsection{Overview}
In this breakout session, comprised of all participants, the conversation was seeded with the following discussion topics: (1) supervised vs. unsupervised learning, (2) transparent vs. black-box methods, (3) probabilistic inference modeling, (4) real-time vs. offline learning, (5) anomaly detection, (6) inference based on missing or noisy data, (7) dimensionality reduction (feature extraction, feature selection), and (7) levels of multimodal data fusion: raw data, feature, decision, etc.

\subsection{Successes, Challenges, and Barriers} 
Recent advances in data analytics and machine learning algorithms have made significant impacts in several domains by revolutionizing the way we process domain-specific data and the techniques uses to evaluate hypotheses or make decisions using these curated data. The field of MMDF has also benefited from these advancements, with methodologies and theories employed in this regard influenced by this fast-paced progress. In particular, the following observations were made: 

\begin{itemize}
\item   \textbf{Supervised vs. Unsupervised Learning:} Nothing is totally unsupervised, and framing the problem already guides the inference model by imposing the model structure and setting the hyper-parameters. Fully supervised models have the huge advantage of learning from the experts (embedded in the labelling information), however they come with several disadvantages including: (1) every supervised decision embeds a series hypothesis (can lead to bias in learning), (2) the high cost of labelling data (e.g., manual labelling of 3D radiology images), (3) the number of experts in different categories is different and is limited (causes unbalanced training detests), and (4) the reality of noisy and unreliable labels from multiple annotators (due to bias and/or disagreements among experts). Given these observation, the panel recommended the use of semi-supervised inference models to capitalize on the advantages of both supervised and semi-supervised learning approache, e.g., using pre-processing steps to generate suggestions for labels (such as taking advantage of natural language processing to find labels from text) or using generative models to produce already labeled datasets. The panel also concluded that moving from fully supervised inference model to unsupervised approaches allows some fields, such as psychology, that have historically been hypothesis-driven to start discovering new insights directly from multimodal data sources.

\item  \textbf{Interperatibility of the Learning Process \& Outcomes:} Deep learning approaches have been rapidly adopted across a wide range of fields because of their accuracy and flexibility \cite{lecun2015deep}. They provide highly scalable solutions for problems in object detection and recognition, machine translation, text-to-speech, and recommendation systems \cite{schmidhuber2015deep}. The majority of deep learning approaches which are based on deep neural networks (DNNs) learn features and labels directly from data. The main challenges here are (1) a requirement for large volumes of labeled data and (2) a difficulty in interpreting the learning process of the algorithm. To address this latter issue while benefiting from the power of DNNs, the panel suggested, for instance, looking into the specific features that are known to have high mutual information with the labels to verify whether latent variables in network have correspondence with these features. 

\item  \textbf{Levels of MMDF:} Classically, the levels at which data fusion would be implemented have been known as: (1) raw data, (2) feature, and (3) decision. In general, raw data fusion is usually done only when data from different modalities are nicely synchronized and have very similar structures (i.e., dimensionality, temporal/spatial resolution). However, in practice, data modalities for most applications have very different structures (e.g., 1D to 4D+ data in patient healthcare information). In these cases, feature level fusion is often employed to represent each modality in a lower dimension where fusion can happen on a smaller dataset. Factor analysis approaches (e.g., Principal Components Analysis (PCA) [12], Exploratory Factor Analysis (EFA), Independent Components Analysis (ICA)) are often used to extract a low dimensional embedding of the original data to facilitate interpretability or computational tractability. Recently, deep learning models such as convolutional neural networks (CNNs) have begun to be utilized as an end-to-end approach that eliminates the need for \emph{a priori} feature extraction, however they remain hindered by the amount of labelled training data that they require. In decision level fusion, the most common in the majority of disciplines due to the parallels to human/system decision making/modeling, the outcomes of individual inference models are integrated to generate a single final decision. As the level of fusion moves from raw to decision (more abstraction), the information content being fused decreases.  It is therefore important to ensure the critical content (higher mutual information with the decision) is passed on, or in cases where it is not obvious what content is critical, this argues for fusion at the raw data or feature level.

\end{itemize}

\subsection{Desired Capabilities and Priorities}
It was confirmed by the panel that among the highest priorities in the field of MMDF is to establish a common scientific framework for MMDF to accelerate solution development by rationalizing an approach to data integration, specifically integrating lessons from MMDF efforts across domains to provide \emph{a priori} recommendations for MMDF methodologies at the project design stage.  The recommendations for next steps in this topic area were (1) to build a common terminology and set of basic principles for MMDF and (2) to design studies to explicitly test the hypothesis that MMDF knowledge gained in one discipline can usefully inform solution design for other disciplines and applications.

\pagebreak
\chapter{Workshop Synthesis:  Roadmap Recommendations and Next Steps}

The workshop discussions led to a consensus on the need for establishing a framework for systematizing the understanding of large and disparate streams of information to produce actionable and reliable decisions in real-time.  This section presents the MMDF Workshop recommendations for a priority roadmap of future cross-disciplinary MMDF research. Based on the major challenges and desired capabilities identified, critical directions are detailed to support MMDF methods, algorithms, and metrics, as outlined below. Addressing these will enable progress toward development and integration of machine learning, statistical methods, signal processing, and optimization with discipline-specific insight for the advancement of MMDF science and technologies.

\vspace{5mm}

\textbf{Question 1:} Can we predict \emph{a priori} for an arbitrary new application where information should optimally be fused (raw data, features, decisions)?  Can we predict \emph{a priori} which architectures will be most effective (decentralized, distributed, hierarchical)?  

\vspace{5mm}

\textbf{Question 2:} How can we take data characteristics and social/economic/physical context into account in the above process?

\vspace{5mm}

\textbf{Question 3:} How can we take system performance demand  (assessment vs. prediction vs. detection vs. inference, etc.) into account in the above process? 

\vspace{5mm}

\textbf{Question 4:} How can we quantitatively evaluate the value (economically or in terms of information content) of the contribution of each modality to the quality of outputs and particularly relative to independent analyses using each modality independently?

\vspace{5mm}

To address these long-range fundamental questions, the workshop attendees identified key critical first steps required to support progress in this field.  Primary among these is \textbf{developing a domain-independent MMDF framework and vocabulary} and \textbf{conducting a comprehensive cross-disciplinary MMDF literature review}.  It is envisioned that this will integrate terminology and concepts utilized in many domains--disciplinary applications in addition to data theory.  In addition there is a critical need for \textbf{a shared evaluation process for recommending and testing MMDF methods} in a more apples-to-apples manner.  To achieve these two goals, the following \textbf{Priority Research Agendas} are proposed:

\begin{itemize}
    \item Synthesizing a domain-independent MMDF vocabulary
    \item Developing a generic architecture framework for MMDF 
    \item Building a cost-benefit evaluation model for MMDF performance 
    \item Defining cross-application MMDF performance metrics
    \item Assembling a taxonomy of applications (relations based on data, application characteristics, or other)
    \item Determining MMDF data quantity and quality needs as a function of generic application characteristics
    \item Facilitating cross-disciplinary transfer of methodologies, algorithms, and evaluation metrics
\end{itemize}

The MMDF2018 workshop achieved an initial approach toward identifying a common vocabulary for the multimodal data fusion concept; a literature review is already underway, with projected publication in Fall 2018.  Moving forward it is envisioned that MMDF research initiatives must be developed to nurture needed cross-disciplinary convergence by (1) explicitly bringing together experts from diverse fields to work together in knowledge discovery and (2) providing discipline-agnostic data-oriented outputs that facilitate broad cross-disciplinary progress in data integration efforts.  

\pagebreak

\chapter{Appendices}
\vspace{-.2in}
\section{Workshop Agenda}
\vspace{-.2in}
\label{sec:agenda}
\vspace{-.1in}
\figProgramone

\figProgramtwo

\figProgramthree

\hfill
\pagebreak
\section{Participant List}
\label{sec:list}

\tabParticipants

\pagebreak
\section{Previous Workshops with Related Themes}
\label{sec:pastworkshops}

\tabpast

\pagebreak
\section{Research Examples in Proposed Framework}
\label{sec:examples}

\tabApplications

{
\bibliographystyle{IEEEbib}
\small
\bibliography{report}
}

\end{document}

%% file: figs.tex
\newcommand{\figLogo}{
\begin{figure}[h!]
  \includegraphics[width=\linewidth,trim=0in 0in 0in 0in, clip=true]{figures/Logo.pdf}
\label{fig:Logo}
\end{figure}
}

\newcommand{\figOverview}{
\begin{figure}[b]
  \centering
  \includegraphics[width=\linewidth,trim=0in 0in 0in 0in, clip=true]{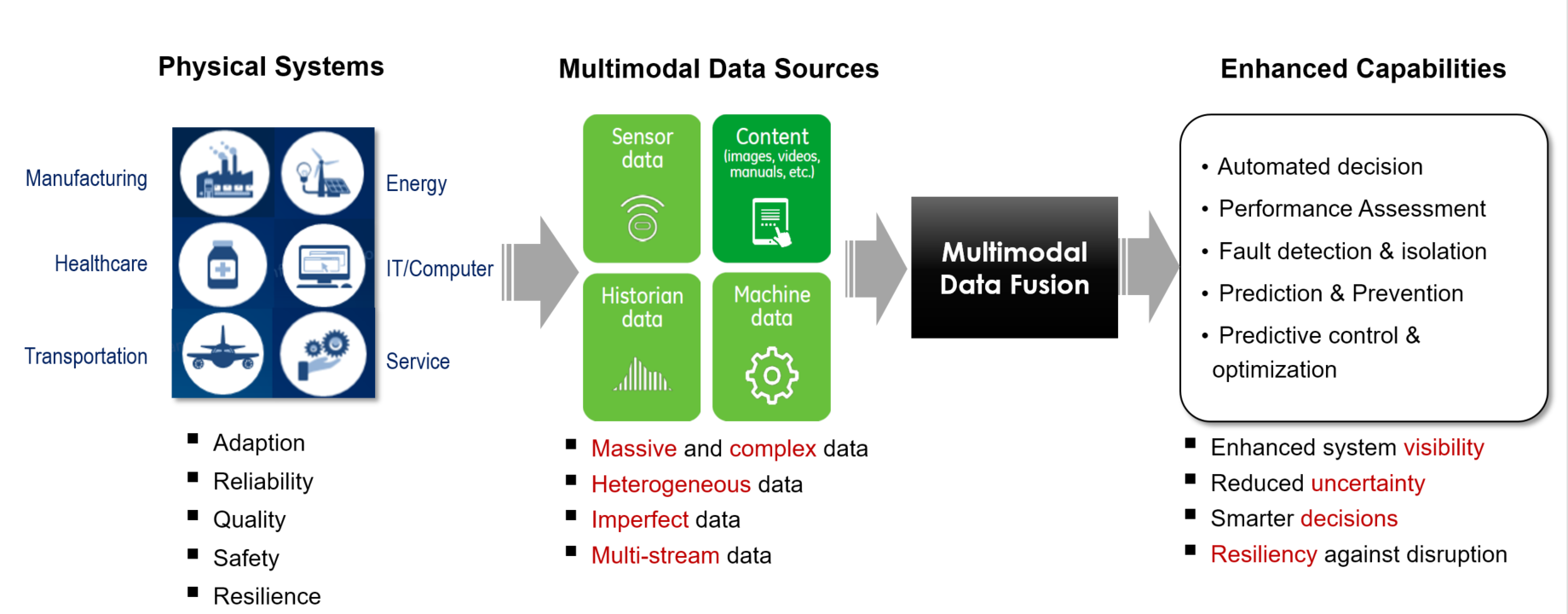}
  \caption{Utility of multimodal data fusion across disciplines and general goals.}
\label{fig:Overview}
\end{figure}
}

\newcommand{\figPlan}{
\begin{figure}[h]
  \centering
  \includegraphics[width=\linewidth,trim=0in 0in 0in 0in, clip=true]{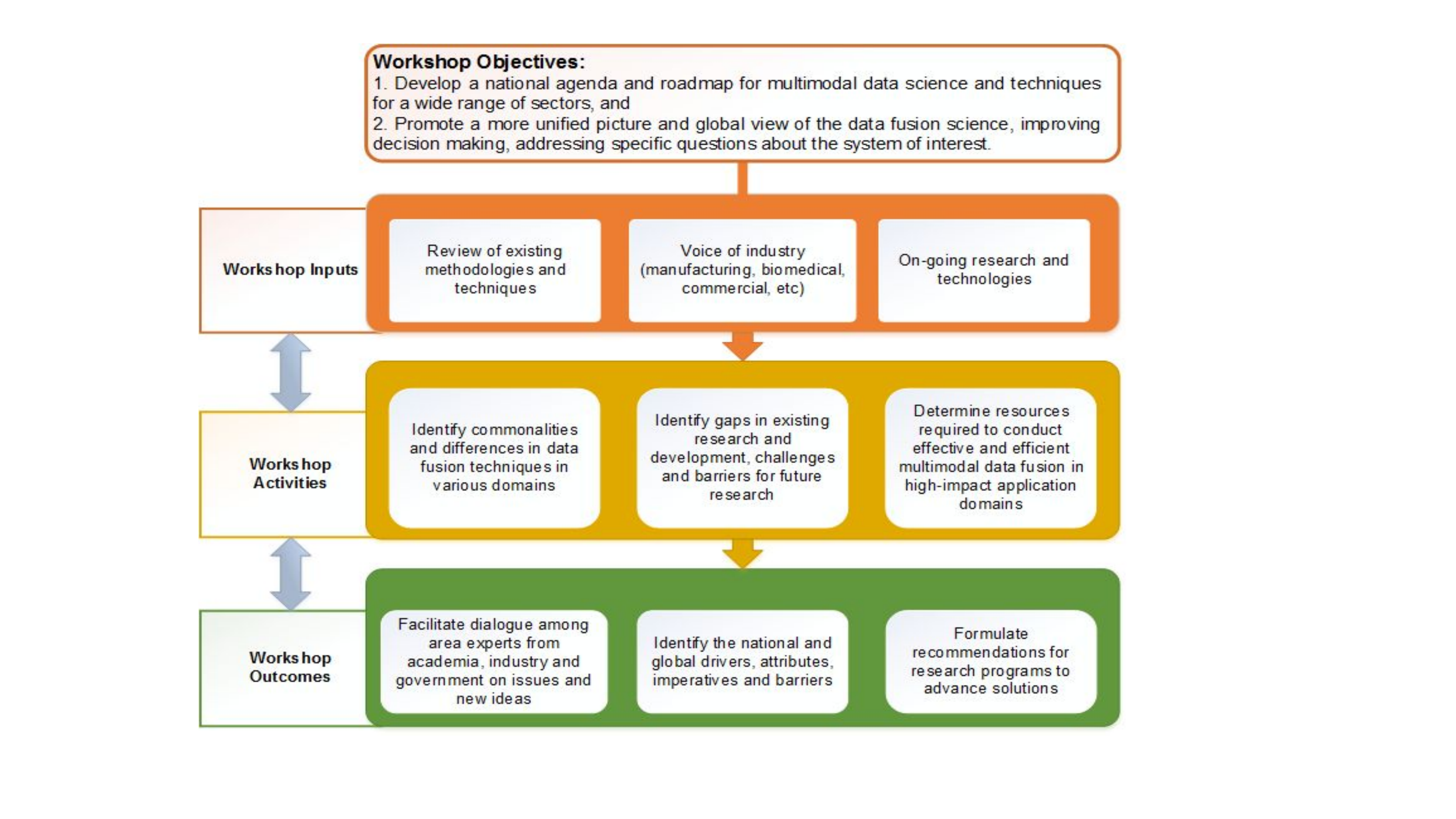}
  \caption{MMDF Workshop Plan.}
\label{fig:Plan}
\end{figure}
}

\newcommand{\figFramework}{
\begin{figure}[h]
  \centering
  \includegraphics[width=.8\linewidth,trim=0in 0in 0in 0in, clip=true]{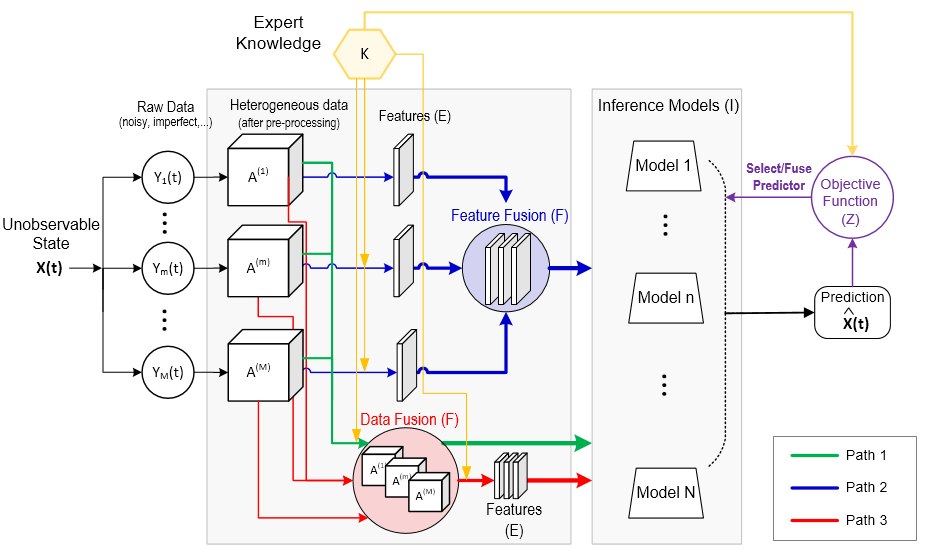}
  \caption{Proposed generalized multimodal data fusion framework to seed discussion.}
\label{fig:Framework}
\end{figure}
}

\newcommand{\figProgramone}{
\begin{figure}[h]
  \centering
  \includegraphics[width=\linewidth,trim=0in 0in 0in 0in, clip=true]{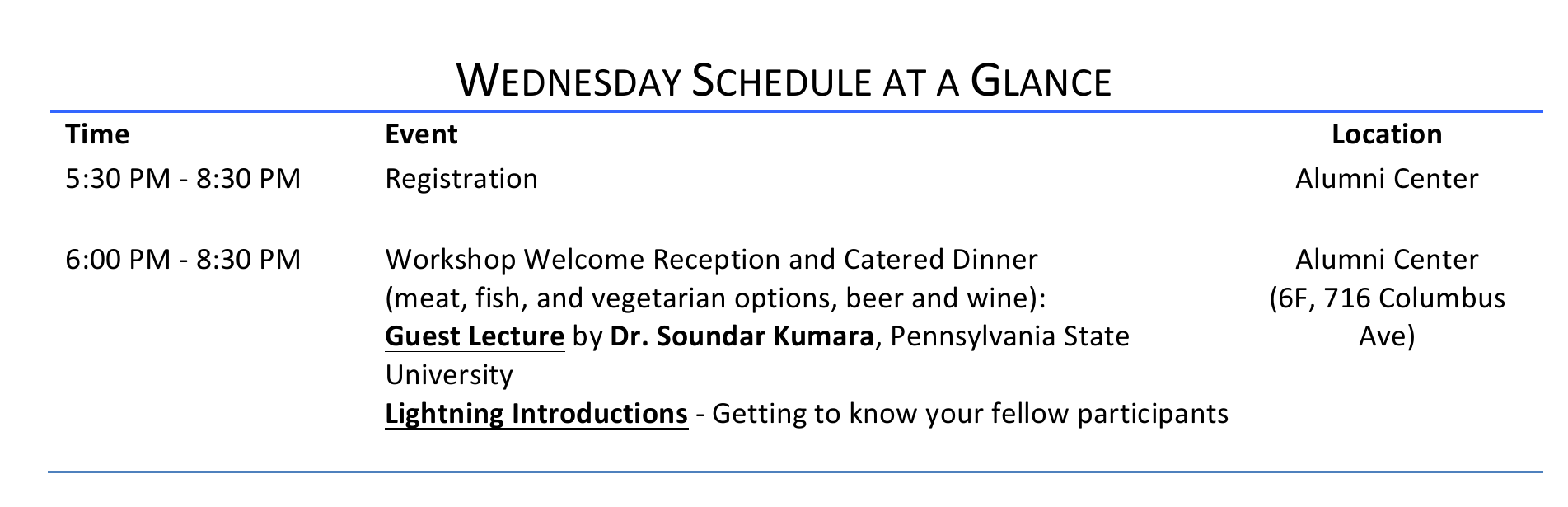}
  \vspace{-.5in}
\label{fig:programone}
\end{figure}
}

\newcommand{\figProgramtwo}{
\begin{figure}[h]
  \centering
  \includegraphics[width=\linewidth,trim=0in 0in 0in 0in, clip=true]{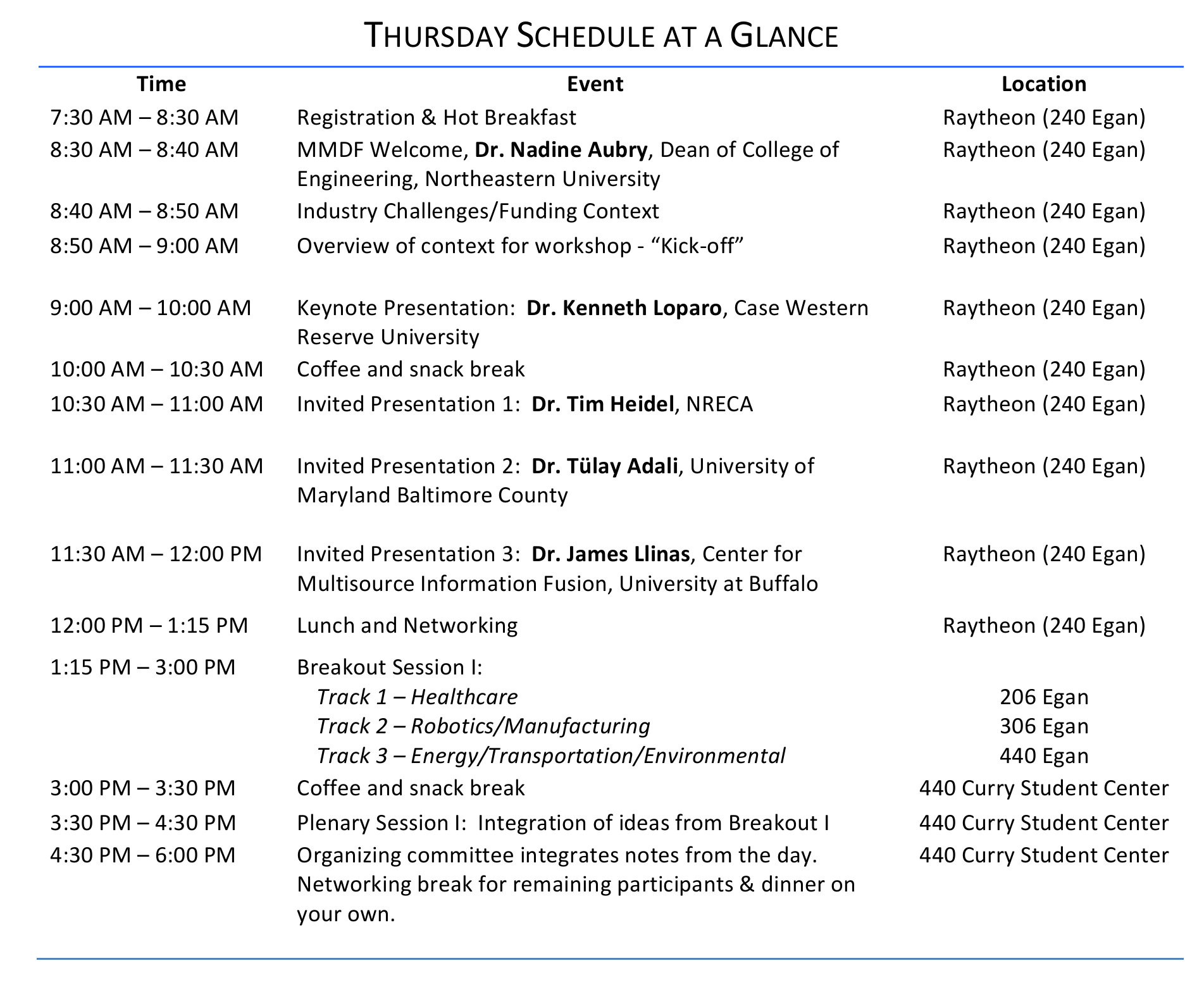}
  \vspace{-.1in}
\label{fig:programtwo}
\end{figure}
}

\newcommand{\figProgramthree}{
\begin{figure}[h]
  \centering
  \includegraphics[width=\linewidth,trim=0in 0in 0in 0in, clip=true]{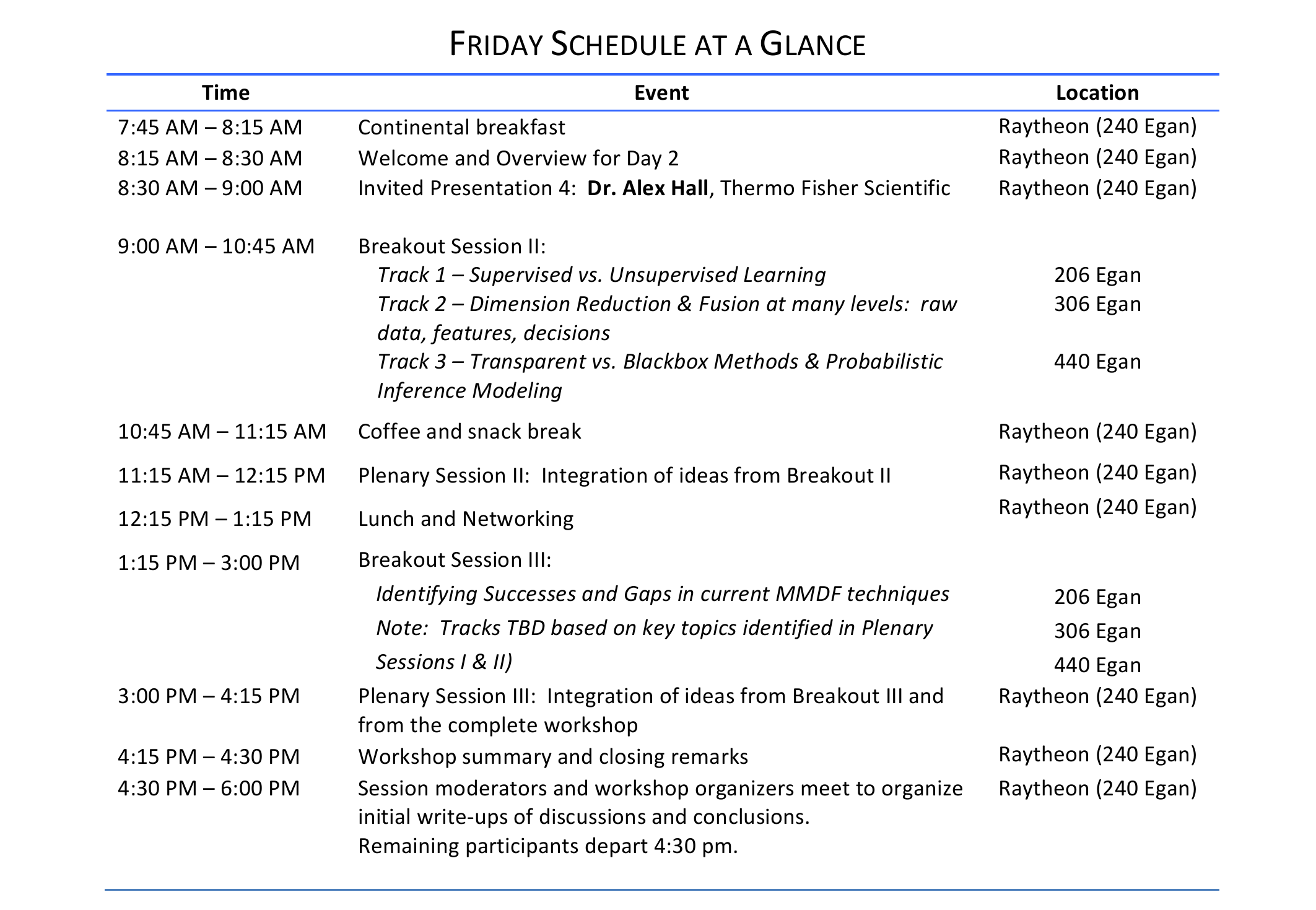}
\label{fig:programthree}
\end{figure}
}

\newcommand{\figParticipants}{
\begin{figure}[h]
  \centering
  \includegraphics[scale=0.7,trim=0in 0in 0in 0in, clip=true]{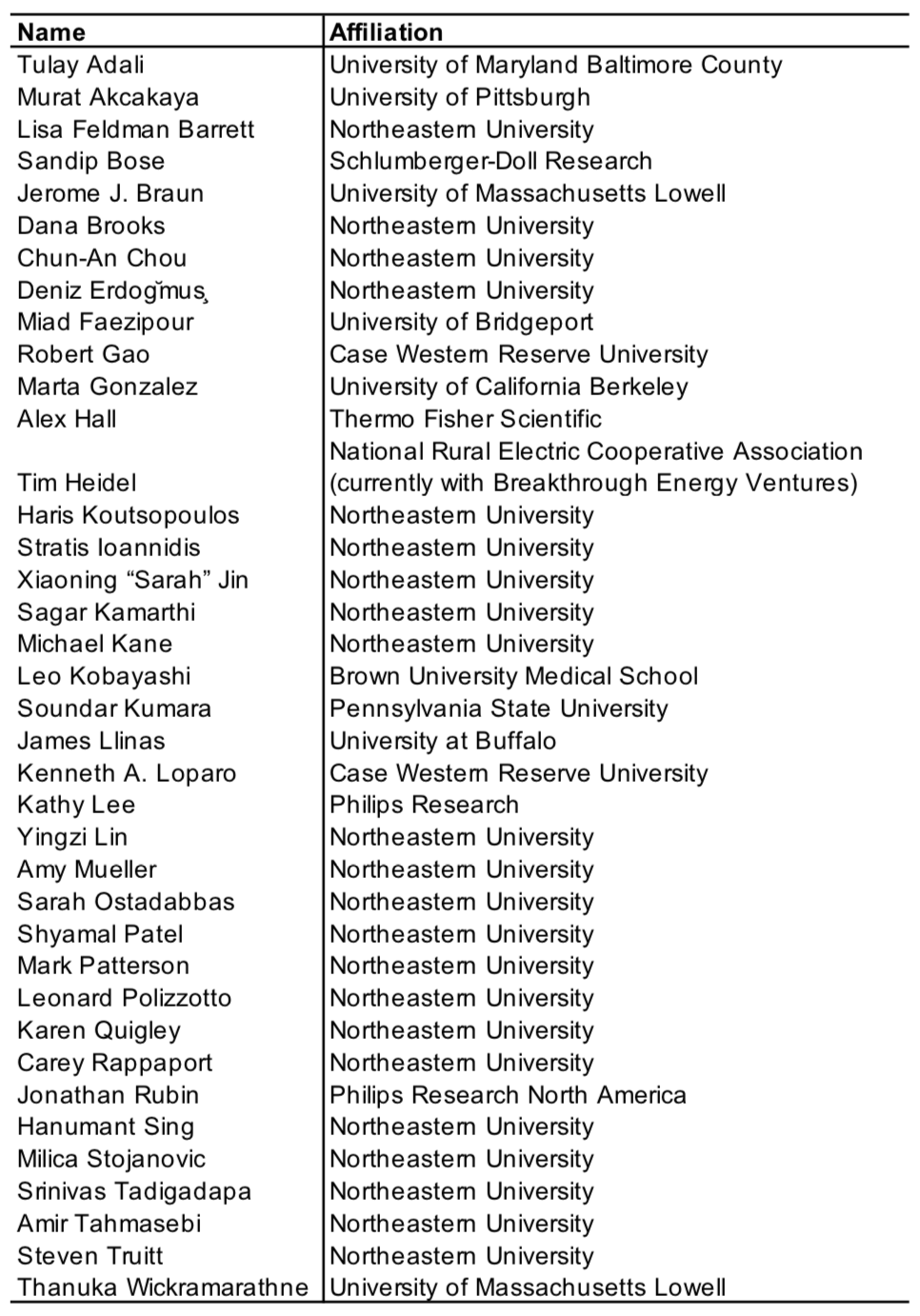}
\label{fig:participants}
\end{figure}
}

\newcommand{\figMfgChallenges}{
\begin{figure}[h]
  \centering
  \includegraphics[scale=0.7,trim=0in 0in 0in 0in, clip=true]{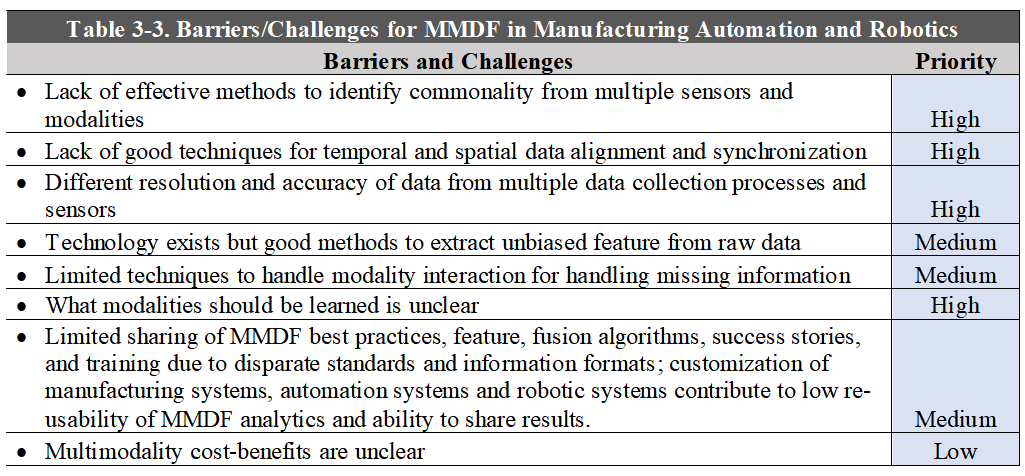}
\label{fig:Mfg}
\end{figure}
}

\newcommand{\figMfgCapabilities}{
\begin{figure}[h]
  \centering
  \includegraphics[scale=0.7,trim=0in 0in 0in 0in, clip=true]{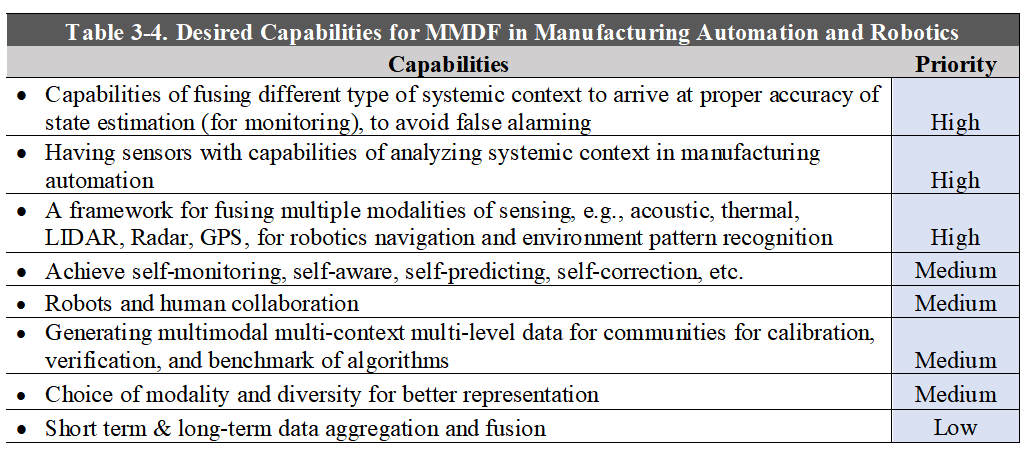}
\label{fig:Mfg}
\end{figure}
}

\newcommand{\figHCCapabilities}{
\begin{figure}[h]
  \centering
  \includegraphics[scale=0.5,trim=0in 0in 0in 0in, clip=true]{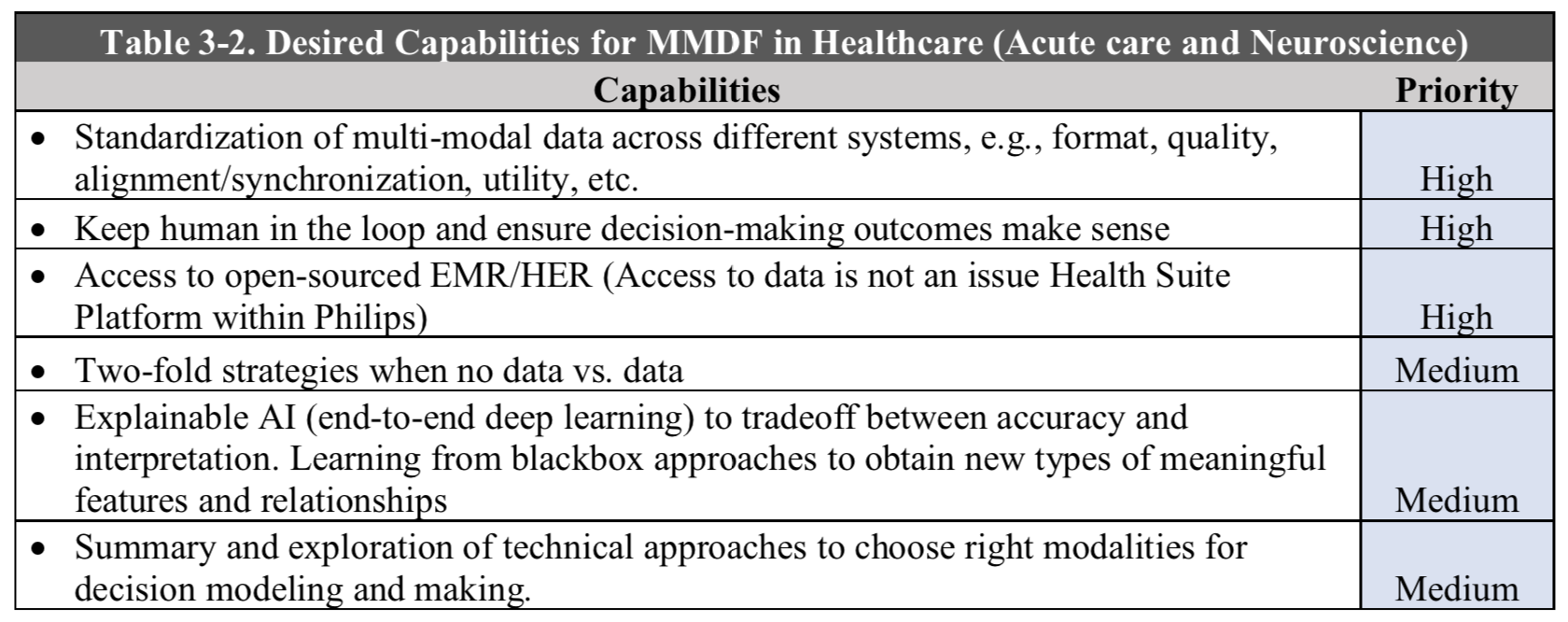}
\label{fig:HC}
\end{figure}
}

\newcommand{\figEnvChallenges}{
\begin{figure}[h]
  \centering
  \includegraphics[scale=0.7,trim=0in 0in 0in 0in, clip=true]{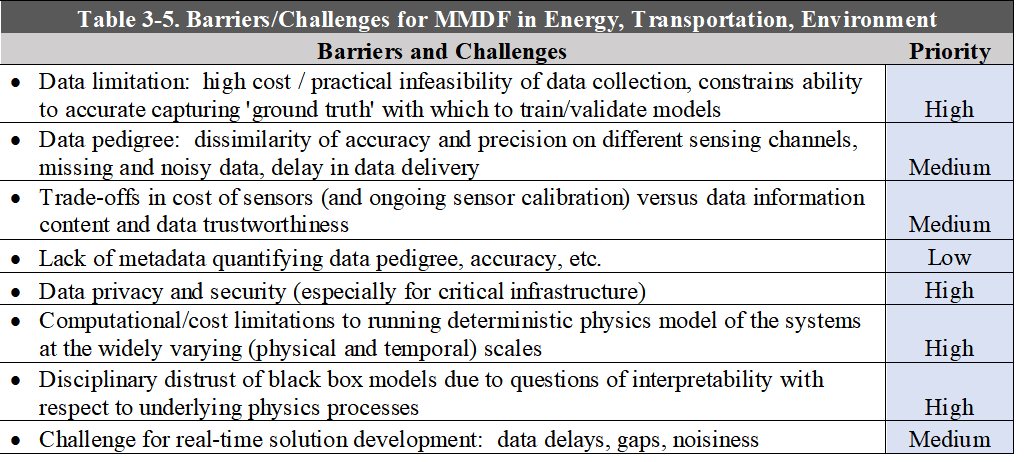}
\label{fig:Mfg}
\end{figure}
}

\newcommand{\figEnvCapabilities}{
\begin{figure}[h]
  \centering
  \includegraphics[scale=0.7,trim=0in 0in 0in 0in, clip=true]{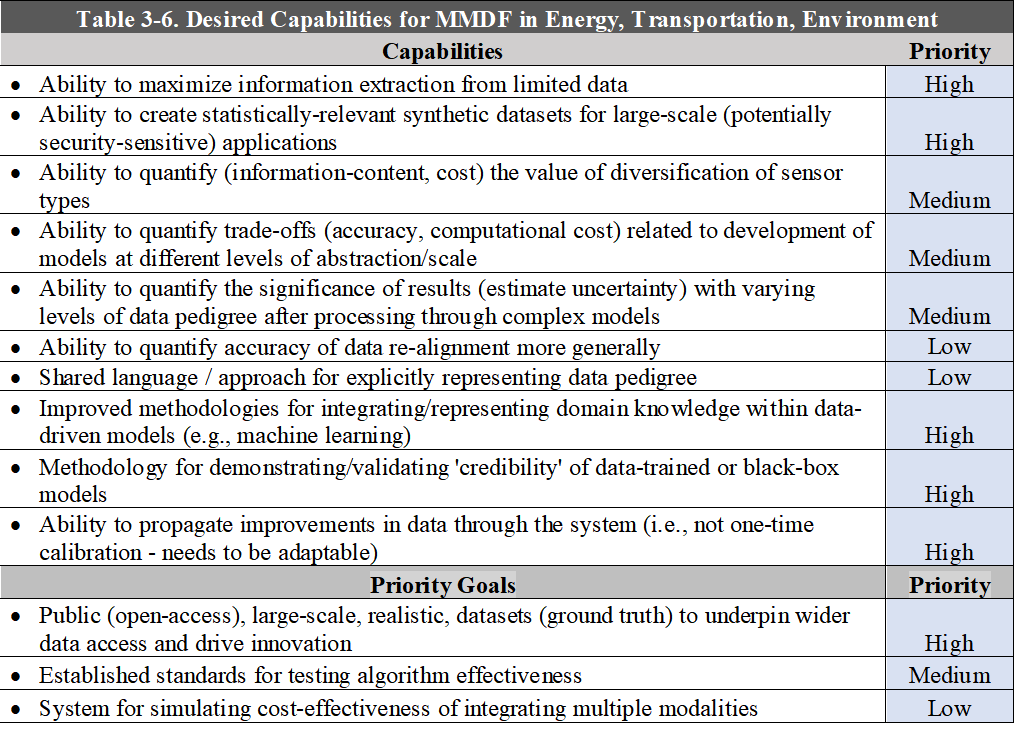}
\label{fig:Mfg}
\end{figure}
}

%% file: tabs.tex
\newcommand{\tabApplications}{
\begin{table}[h]
\caption{Illustrative applications mapped onto the generalized framework in \figref{Framework}}
\begin{center}
\includegraphics[width=\linewidth,trim=0in 0in 0in 0in, clip=true]{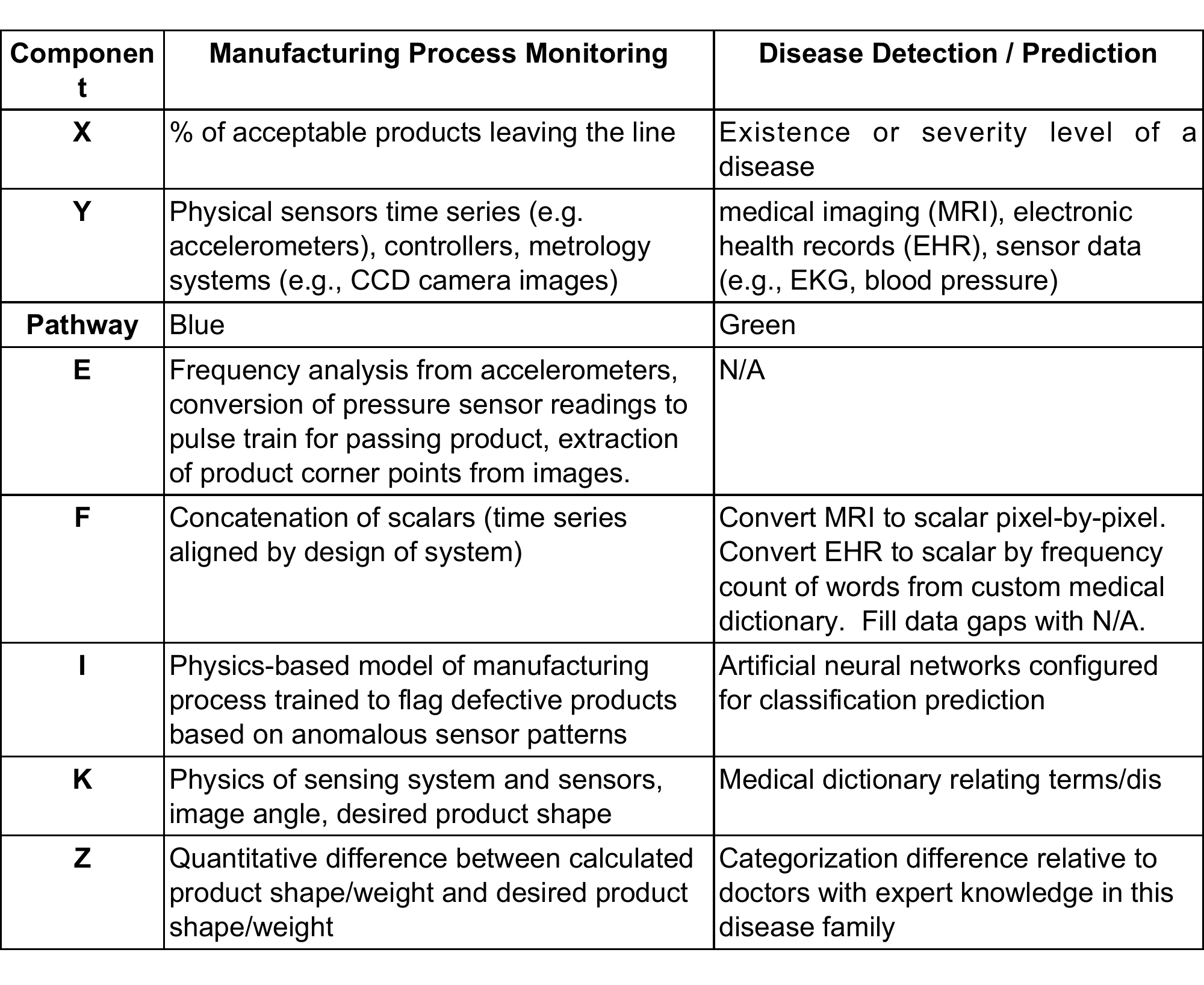}
\label{tbl:Applications}
\end{center}
\end{table}
}

\newcommand{\tabPastWorkshops}{
\begin{table}
\caption{Past workshops with related themes (\com{the references needs to be updated})}
\begin{center}
\includegraphics[width=\linewidth,trim=0in 0in 0in 0in, clip=true]{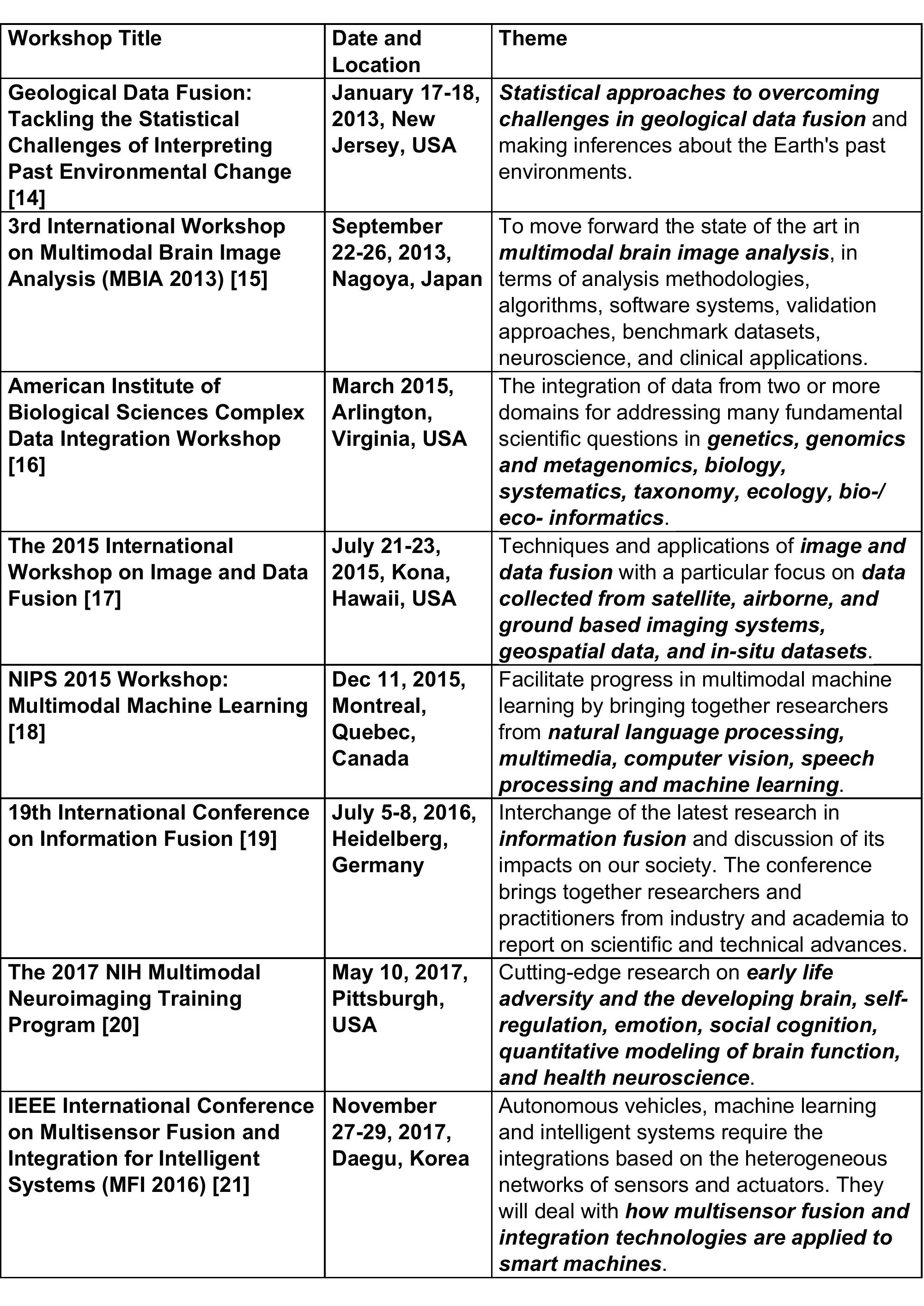}
\label{tbl:PastWorkshops}
\end{center}
\end{table}
}

\newcommand{\tabParticipants}{
\begin{table}[h]
 \vspace{-.2in}
\small
\begin{tabular}{| m{4cm} | m{8cm}|}
\hline
\textbf{Name}                  & \textbf{Affiliation}                         \\
\hline
T{\"u}lay       Adali           & University of Maryland Baltimore County   \\
\hline
Murat         Akcakaya        & University of Pittsburgh                    \\
\hline
MD Noor       Alam            & Northeastern University                     \\
\hline
Yaakov        Bar-Shalom      & University Connecticut                         \\
\hline
Alexandre     Bayen           & University of California Berkeley           \\
\hline
Sandip        Bose            & Schlumberger-Doll Research Center \\
\hline
Jerome        Braun           & University of Massachusetts Lowell  \\
\hline
Dana          Brooks          & Northeastern University                 \\
\hline
Viveck        Cadambe         & Pennsylvania State University             \\
\hline
Chun-An       Chou            & Northeastern University                 \\
\hline
Pau           Closas          & Northeastern University                     \\
\hline
Claire        Duggan          & Northeastern University                  \\
\hline
Deniz         Erdogmus        & Northeastern University                 \\
\hline
Miad          Faezipour       & Bridgeport University\\
\hline
Lisa          Feldman Barrett & Northeastern University         \\
\hline
Robert        X. Gao          & Case Western Reserve University   \\
\hline
Marta         Gonzalez        & University of California Berkeley       \\
\hline
Alex          Hall            & ThermoFisher Scientific \\
\hline
Timothy       Heidel          & National Rural Electric Cooperative Association\\
\hline
Stratis       Ioannidis       & Northeastern University             \\
\hline
Xiaoning      Jin             & Northeastern University                 \\
\hline
Mark F.       Kahn            & LM space systems         \\     
\hline
Sagar         Kamarthi        & Northeastern University                 \\
\hline
Michael       Kane            & Northeastern University                    \\
\hline
Leo           Kobayashi       & Brown University \\
\hline
Haris         Koutsopoulos    & Northeastern University         \\
\hline
Soundar       Kumara          & Pennsylvania State University        \\
\hline
Kathy         Lee             & Philips Research North America \\
\hline
Yingzi        Lin             & Northeastern University         \\
\hline
James         Llinas          & University of Buffalo                      \\
\hline
Kenneth       Loparo          & Case Western Reserve University               \\
\hline
Amy           Mueller         & Northeastern University         \\
\hline
Sarah         Ostadabbas      & Northeastern University                 \\
\hline
Shyamal       Patel           & Pfizer Innovation Research (PfIRe) Lab         \\
\hline
Mark          Patterson       & Northeastern University     \\
\hline
Leonard       Polizzotto      & Northeastern University                 \\
\hline
Karen         Quigley         & Northeastern University         \\
\hline
Carey         Rappaport       & Northeastern University     \\
\hline
Jonathan      Rubin           & Philips Research North America \\
\hline
Hanumant      Singh           & Northeastern University         \\
\hline
Milica        Stojanovic      & Northeastern University         \\
\hline
Amir           Tahmasebi      & Philips Research North America \\
\hline
Steven        Truitt          & MITRE \\
\hline
Ryan          Wang            & Northeastern University     \\
\hline
Thanuka       Wickramarathne  & University of Massachusetts Lowell \\
\hline
Smaine        Zeroug          & Schlumberger-Doll Research Center\\
\hline
\end{tabular}
  \vspace{-.2in}
\end{table}
}

\newcommand{\tabHCChallenges}{
\begin{table}[h]
\caption{Barriers/Challenges for MMDF in Healthcare (Acute care and Neuroscience).}
  \vspace{-.1in}
\begin{center}
\includegraphics[width=.9\linewidth,trim=0in 0in 0in 0in, clip=true]{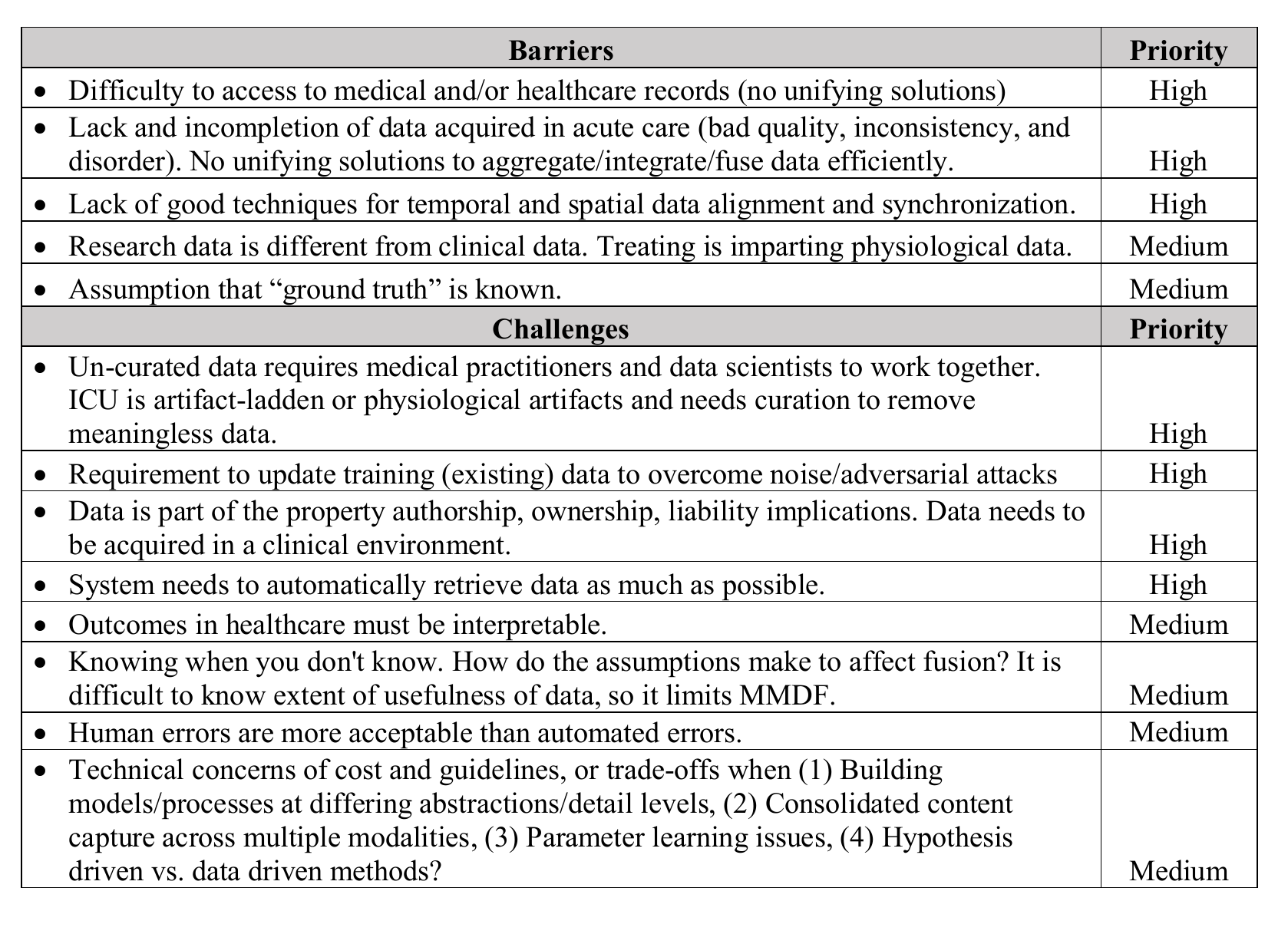}
\label{tbl:HCChallenges}
  \vspace{-.1in}
\end{center}
\end{table}
}

\newcommand{\tabHCCapabilities}{
\begin{table}[h]
\caption{Desired Capabilities for MMDF in Healthcare (Acute care and Neuroscience).}
  \vspace{-.1in}
\begin{center}
\includegraphics[width=.9\linewidth,trim=0in 0in 0in 0in, clip=true]{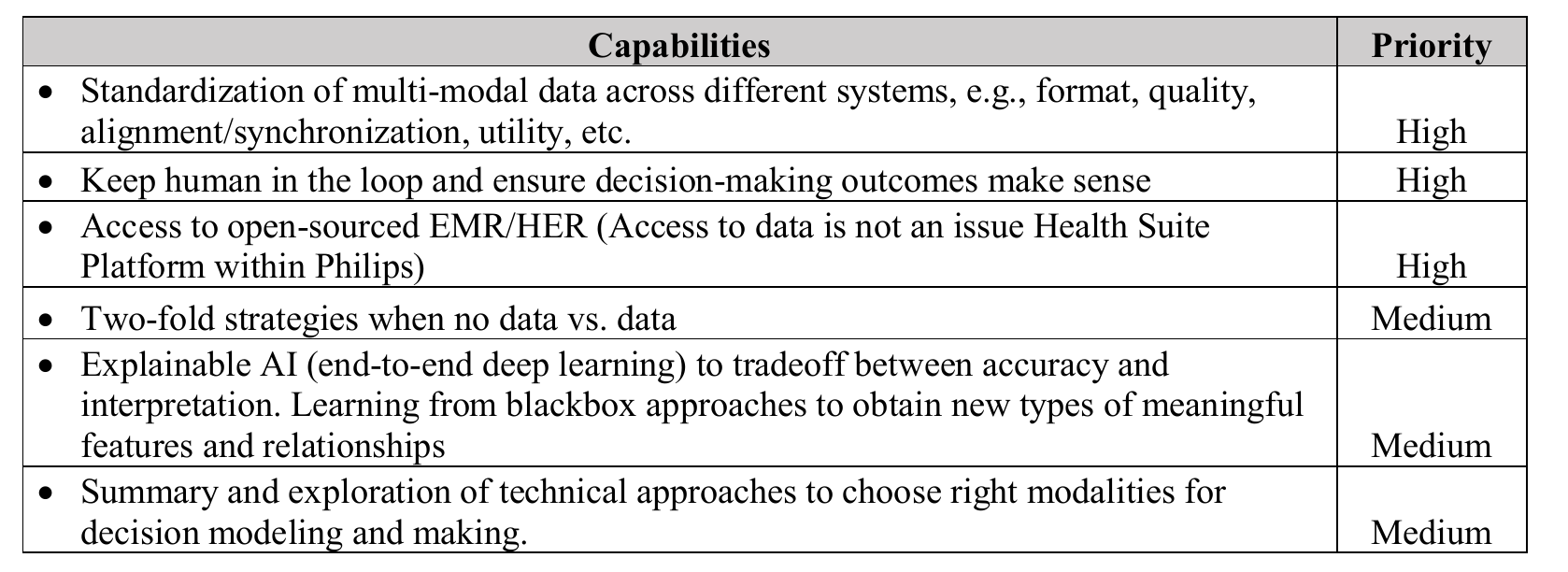}
\label{tbl:HCCapabilities}
  \vspace{-.1in}
\end{center}
\end{table}
}

\newcommand{\tabMfgChallenges}{
\begin{table}[h]
\caption{Barriers/Challenges for MMDF in Manufacturing Automation and Robotics.}
  \vspace{-.1in}
\begin{center}
\includegraphics[width=.9\linewidth,trim=0in 0in 0in 0in, clip=true]{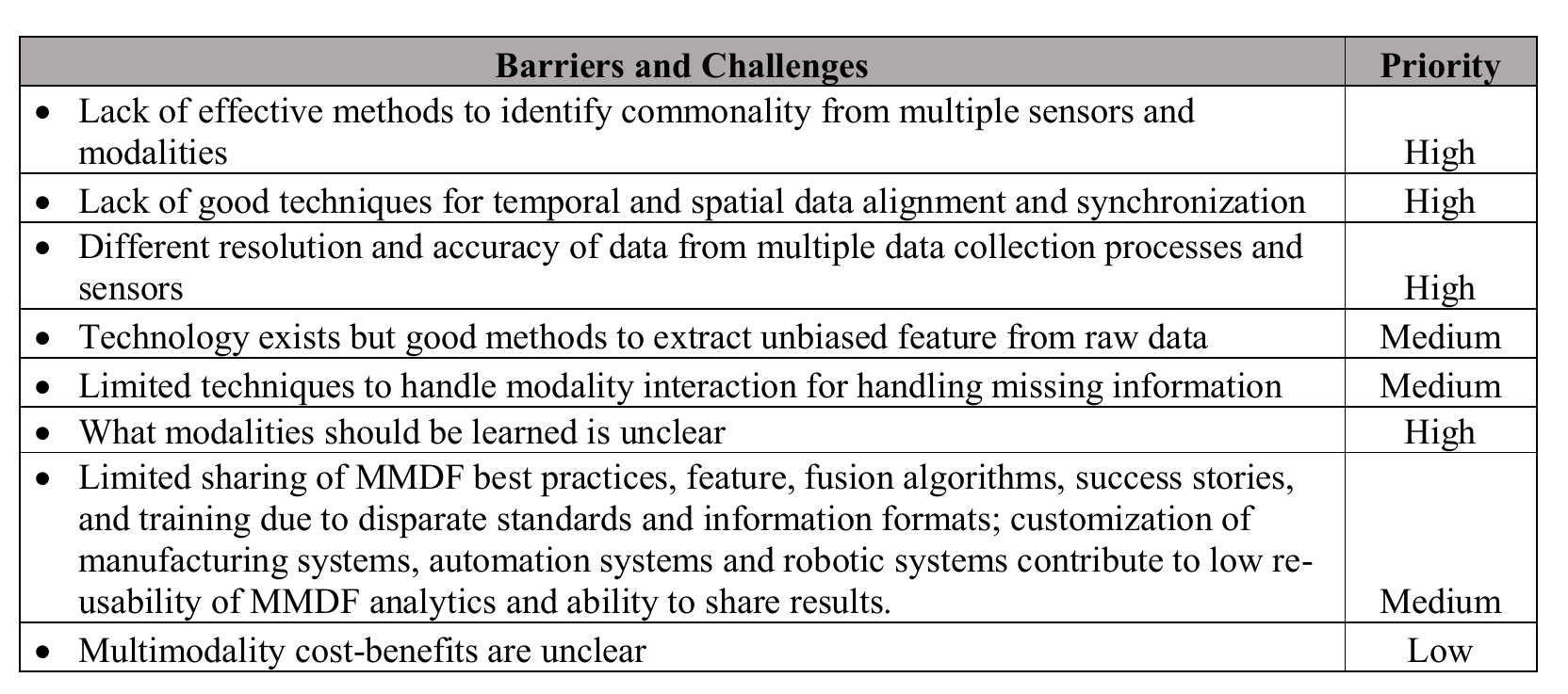}
\label{tbl:MfgChallenges}
  \vspace{-.1in}
\end{center}
\end{table}
}

\newcommand{\tabMfgCapabilities}{
\begin{table}[h]
\caption{Desired Capabilities for MMDF in Manufacturing Automation and Robotics.}
  \vspace{-.1in}
\begin{center}
\includegraphics[width=.9\linewidth,trim=0in 0in 0in 0in, clip=true]{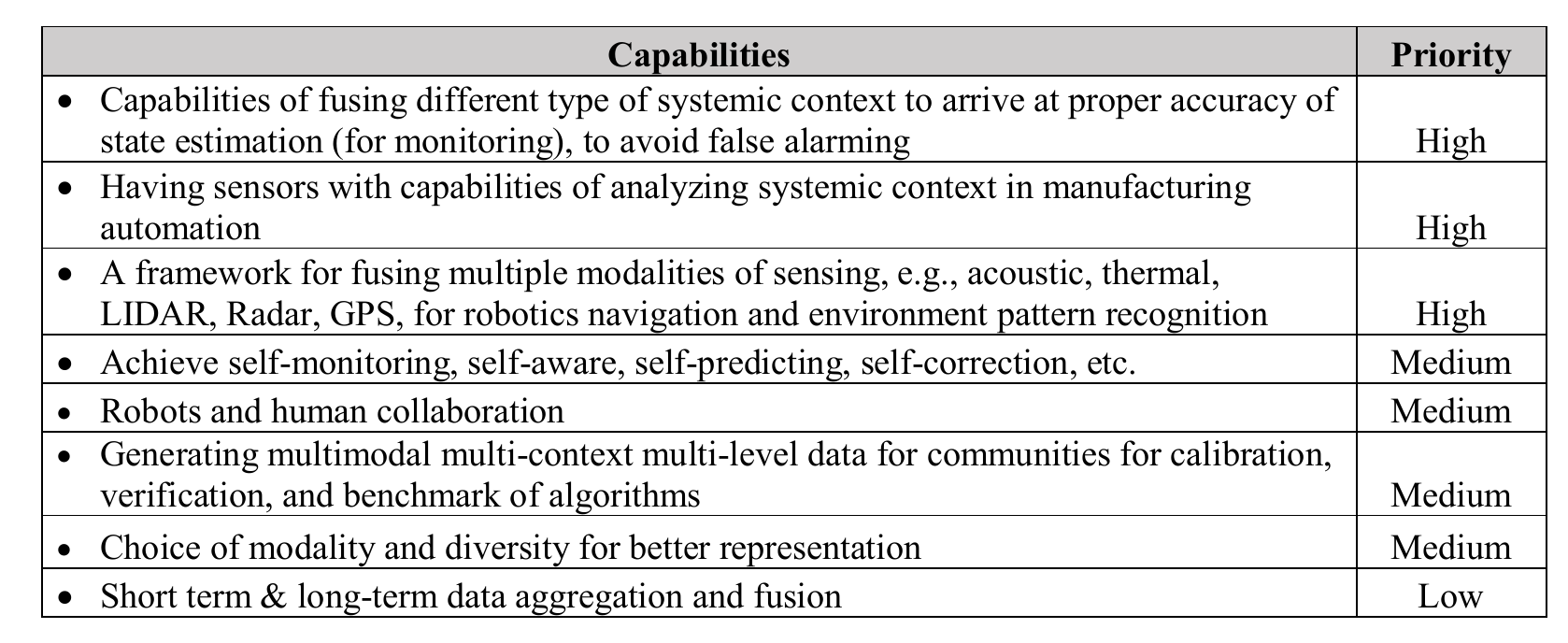}
\label{tbl:MfgCapabilities}
  \vspace{-.1in}
\end{center}
\end{table}
}

\newcommand{\tabEnvChallenges}{
\begin{table}[h]
\caption{Barriers/Challenges for MMDF in Energy, Transportation, Environment.}
  \vspace{-.1in}
\begin{center}
\includegraphics[width=.9\linewidth,trim=0in 0in 0in 0in, clip=true]{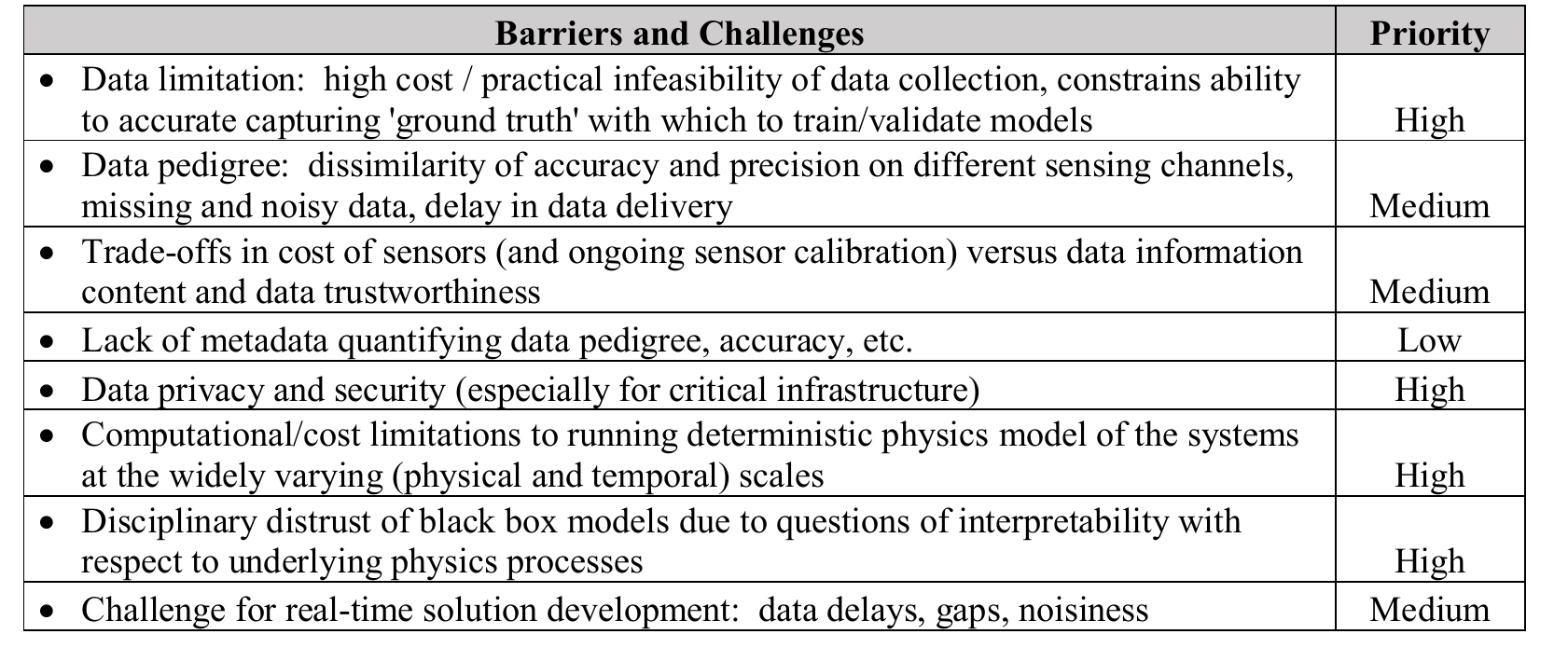}
\label{tbl:EnvChallenges}
  \vspace{-.1in}
\end{center}
\end{table}
}

\newcommand{\tabEnvCapabilities}{
\begin{table}[h]
  \vspace{-.1in}
\caption{Desired Capabilities for MMDF in Energy, Transportation, Environment.}
  \vspace{-.1in}
\begin{center}
\includegraphics[width=.9\linewidth,trim=0in 0in 0in 0in, clip=true]{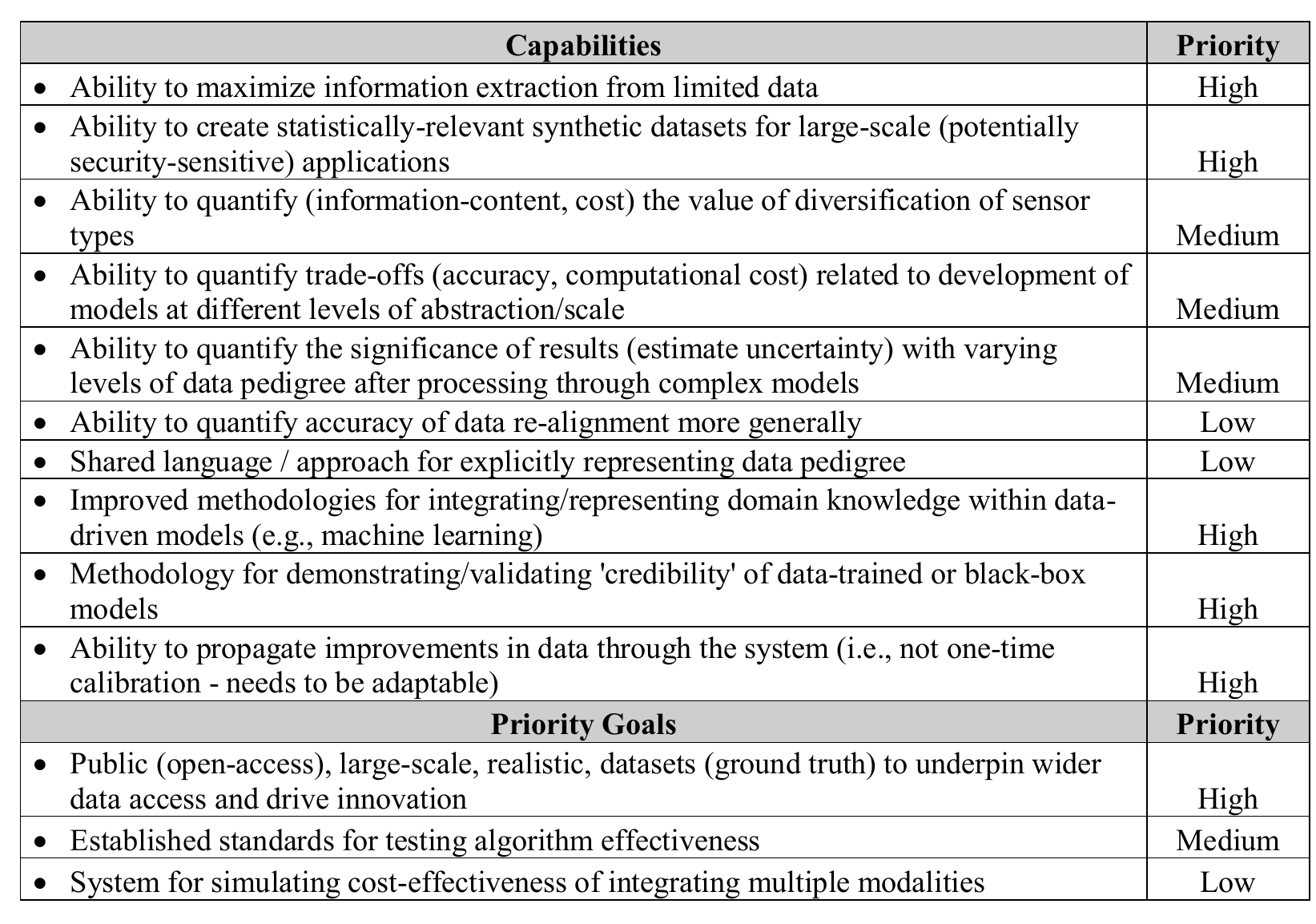}
\label{tbl:EnvCapabilities}
  \vspace{-.1in}
\end{center}
\end{table}
}

\newcommand{\tabpast}{
\begin{tabular} { | m{4cm} | m{3cm}| m{7cm} |}
\hline
\textbf{Workshop Title} &    \textbf{Location and Date}  &      \textbf{Theme} \\

\hline
Geological Data Fusion: Tackling the Statistical Challenges of Interpreting Past Environmental Change \cite{workshop1} & January 17-18, 2013, New Jersey, USA & Statistical approaches to overcoming challenges in geological data fusion and making inferences about the Earth's past environments.\\

\hline
3rd International Workshop on Multimodal Brain Image Analysis (MBIA 2013) \cite{workshop2} & September 22-26, 2013, Nagoya, Japan & To move forward the state of the art in multimodal brain image analysis, in terms of analysis methodologies, algorithms, software systems, validation approaches, benchmark datasets, neuroscience, and clinical applications.\\

\hline
American Institute of Biological Sciences Complex Data Integration Workshop \cite{workshop3} & March 2015, Arlington, Virginia, USA & The integration of data from two or more domains for addressing many fundamental scientific questions in genetics, genomics and metagenomics, biology, systematics, taxonomy, ecology, bio-/eco- informatics. \\

\hline
The 2015 International Workshop on Image and Data Fusion \cite{workshop4}  & July 21-23, 2015, Kona, Hawaii, USA & Techniques and applications of image and data fusion with a particular focus on data collected from satellite, airborne, and ground based imaging systems, geospatial data, and in-situ datasets.\\

\hline
NIPS 2015 Workshop: Multimodal Machine Learning \cite{workshop5} & Dec 11, 2015, Montreal, Quebec, Canada & Facilitate progress in multimodal machine learning by bringing together researchers from natural language processing, multimedia, computer vision, speech processing and machine learning. \\

\hline
19th International Conference on Information Fusion \cite{workshop6} & July 5-8, 2016, Heidelberg, Germany & Interchange of the latest research in information fusion and discussion of its impacts on our society. The conference brings together researchers and practitioners from industry and academia to report on scientific and technical advances. \\

\hline
The 2017 NIH Multimodal Neuroimaging Training Program \cite{workshop7} & May 10, 2017, Pittsburgh, USA & Cutting-edge research on early life adversity and the developing brain, self-regulation, emotion, social cognition, quantitative modeling of brain function, and health neuroscience.\\

\hline
IEEE International Conference on Multisensor Fusion and Integration for Intelligent Systems (MFI 2016) \cite{workshop8} & November 27-29, 2017, Daegu, Korea & Autonomous vehicles, machine learning and intelligent systems require the integrations based on the heterogeneous networks of sensors and actuators. They will deal with how multisensor fusion and integration technologies are applied to smart machines.\\

\hline
\end{tabular}
}